\documentclass[apj]{emulateapj}

\submitted{\it Accepted for publication in the Astrophysical Journal}

\usepackage{apjfonts}
\usepackage{natbib}
\usepackage{amsfonts}
\usepackage{amsmath}
\usepackage{graphicx}
\usepackage{xspace}
\usepackage{hyperref}

\newcommand{\code}[1]{\texttt{#1}}
\newcommand\mesa{\code{MESA}\xspace}
\newcommand\emcee{\code{emcee}\xspace}
\newcommand{\me}{\ensuremath M_{\rm E}}
\newcommand{\mj}{\ensuremath M_{\rm J}}
\newcommand{\rj}{\ensuremath R_{\rm J}}
\newcommand{\mc}{\ensuremath M_{\rm c}}
\newcommand\grad\nabla
\newcommand\grada{\nabla_{\rm ad}}
\newcommand\gradmu{\nabla_\mu}
\newcommand\gradr{\nabla_{\rm rad}}
\newcommand\B{\frac{\varphi}{\delta}\gradmu}
\newcommand\teff{T_{\rm eff}}
\newcommand\teq{T_{\rm eq}}
\newcommand\yatm{Y_{\rm atm}}
\newcommand\rvol{R_{\rm vol}}
\newcommand\req{R_{\rm eq}}
\newcommand\rpol{R_{\rm pol}}
\newcommand\tdmx{T_{\rm phase}}
\newcommand\dtdmx{\Delta\tdmx}
\newcommand\xhe{x_{\rm He}}
\renewcommand\bv{Br\"unt-V\"ais\"al\"a\xspace}
\newcommand\pr{\rm Pr}
\newcommand\rcrit{R_{\rm crit}}
\newcommand\scvhi{SCvH-I\xspace}
\renewcommand\r{R_\rho}
\newcommand\rhoc{\rho_{\rm c}}
\newcommand\mzenv{M_{Z,{\rm env}}}
\newcommand\mz{M_Z}

\begin{document}

\shortauthors{Mankovich, Fortney and Moore}
\shorttitle{Helium Rain in Jupiter}
\title{Bayesian Evolution Models for Jupiter with Helium Rain \\and Double-diffusive Convection}

\author{Christopher Mankovich}
\email{\tt cmankovich@ucsc.edu}
\author{Jonathan J. Fortney}
\author{Kevin L. Moore}
\affil{Department of Astronomy and Astrophysics, University of California, Santa Cruz, CA 95064, USA} 

\begin{abstract}
Hydrogen and helium demix when sufficiently cool, and this bears on the evolution of all giant planets at large separations at or below roughly a Jupiter mass. We model the thermal evolution of Jupiter, including its evolving helium distribution following results of ab initio simulations for helium immiscibility in metallic hydrogen. After 4 Gyr of homogeneous evolution, differentiation establishes a thin helium gradient below 1 Mbar that dynamically stabilizes the fluid to convection. The region undergoes overstable double-diffusive convection (ODDC), whose weak heat transport maintains a superadiabatic temperature gradient. With a generic parameterization for the ODDC efficiency, the models can reconcile Jupiter's intrinsix flux, atmospheric helium content, and radius at the age of the solar system if the Lorenzen et al. H-He phase diagram is translated to lower temperatures. We cast the evolutionary models in an MCMC framework to explore tens of thousands of evolutionary sequences, retrieving probability distributions for the total heavy element mass, the superadiabaticity of the temperature gradient due to ODDC, and the phase diagram perturbation. The adopted \scvhi equation of state favors inefficient ODDC such that a thermal boundary layer is formed, allowing the molecular envelope to cool rapidly while the deeper interior actually heats up over time. If the overall cooling time is modulated with an additional free parameter to imitate the effect of a colder or warmer EOS, the models favor those that are colder than \scvhi. In this case the superadiabaticity is modest and a warming or cooling deep interior are equally likely.
\end{abstract}


\section{Introduction}\label{s.intro}
Cool giant planets are relics of the protoplanetary systems from which they formed in the sense that they do not fuse protons, and they are well-bound enough that even hydrogen does not escape appreciably over tens of billions of years.  Their thermal evolution is thus relatively simple, and understanding it empowers us to use the present states of giant planets to learn about their history and formation. The open questions about planet formation thus motivate a comprehensive theory of giant planet evolution, which will continue to be driven heavily by our own, well-studied giant planets, Jupiter and Saturn.

A Henyey-type stellar evolution calculation for a Jupiter-mass object was first performed by \cite{graboske1975}, who showed that a convective, homogeneous sphere of fluid hydrogen and helium could cool to Jupiter's observed luminosity over roughly the right timescale, and noted that among all model inputs, the equation of state (EOS) and superadiabaticity of the temperature gradient have the strongest influence on the overall cooling time.  

These two fundamental physical inputs are closely related. The EOS (paired with a hydrostatic model) is necessary to translate the planet's tangible properties (surface temperature and composition; external gravity field, size and shape) into an interior density distribution. Knowledge of the thermodynamic state of matter in these regimes includes understanding any phase transitions that can operate in a Jovian-mass planet's interior, the two most important of which are (i) the transition from molecular hydrogen to its denser, pressure-ionized ``liquid metallic'' phase, and (ii) the limited solubility of neutral helium in that liquid metallic hydrogen once it cools below a critical temperature \citep{1975PhRvB..12.3999S}.  The latter of these two effects has observable ramifications because the helium-rich phase tends to sink, releasing gravitational energy (constituting a power source beyond mere contraction) and depleting the outer envelope in helium \citep{1973ApJ...181L..83S,ss77b}. Ultimately a robust theory of giant planet evolution must reconcile the atmospheric helium mass fraction $\yatm$ with the helium content of the protosolar nebula, and this demand constrains the plausible EOS and H-He phase diagram.

Since the critical temperature for H-He phase separation increases with pressure more slowly than the temperature along a planetary adiabat, the equilibrium helium abundance increases toward the center the planet.  Thus in the limit that the hydrogen-helium mixing ratio is equal to its equilibrium value throughout the liquid metallic hydrogen part of the mantle, there exists a stabilizing helium gradient that acts to mitigate the convectively unstable temperature gradient. In this case the dynamics of the fluid (and the degree of macroscopic vertical heat transport that ensues) are dictated by the competing microscopic diffusion of heat and solute; the fluid is in the double-diffusive regime. In such a region the temperature gradient can be significantly larger than the adiabatic gradient, leading to potentially dramatic modifications to the planet's cooling time. For example, double-diffusive convection has been invoked in recent years to explain Saturn's luminosity excess (\citealt{lc13}; the case of a global heavy-element gradient), the inflation of hot Jupiters \citep{2015ApJ...815...78K}, and Jupiter's late thermal evolution including helium rain \citep{2015MNRAS.447.3422N}, which we are revisiting in this paper. Although differentiation alone contributes additional luminosity, extending the overall cooling time, any superadiabatic temperature structure associated with double-diffusive convection generally cools the surface more quickly. For models undergoing helium rain, cases with adiabatic $P-T$ profiles thus give an upper limit to the cooling time (\citealt{2003Icar..164..228F}, \citealt{2016Icar..267..323P}). The inclusion of double-diffusive convection offers a continuum of shorter cooling times, modulated by the efficiency of the heat transport through the double-diffusive region.

\cite{2015MNRAS.447.3422N} sought a solution for Jupiter's evolution to its current state assuming a superadiabatic temperature profile in the framework of layered double-diffusive convection (LDDC; \citealt{2012ApJ...750...61M}, \citealt{2013ApJ...768..157W}) and found that a suitable combination of LDD layer height and modifications to the H-He phase diagram could match Jupiter's observed 1-bar temperature and $\yatm$. However, it is likely that a quasi-stable turbulent state like the layered structures characterized by the direct hydrodynamics simulations of \cite{2013ApJ...768..157W} would look quite different in the presence of a phase transition and rainout of a main component.  For example, in the context of helium phase separation, homogeneous layers themselves---finite volumes of $(P,\ T)$ space, at effectively uniform helium abundance---are intrinsically unstable to H-He phase separation, and the influence that droplet formation and rainout has on the merging or bifurcation of convective layers, or the transport of solute between layers, has yet to be assessed from the hydrodynamical perspective.  The present work thus relaxes the assumption of layered convection, opting instead to treat the superadiabaticity with a generic parameterization, the only physical content of which is the demand that the temperature gradient lie somewhere between the minimum value for overstable double-diffusion and the upper limit imposed by the Ledoux criterion. This amounts to the criterion that gravity waves be linearly overstable, so that thermal transport is enhanced relative to the purely diffusive case by some degree of turbulence.

Despite growing confidence that helium has begun differentiating in Jupiter's recent past (and billions of years ago in the case of Saturn; see \citealt{2003Icar..164..228F} and \citealt{2016Icar..267..323P}), it is not known whether helium rain alone can resolve the gaps in our understanding of giant planet evolution given Jupiter and Saturn's luminosities and the helium content of their molecular envelopes at the present day. Although the thermodynamic conditions for phase separation of helium from liquid metallic hydrogen have been evaluated since the early work of \cite{1975PhRvB..12.3999S} and \cite{1977PhRvB..15.1914S}, quantitative knowledge covering the relevant pressures and H-He mixing ratios has only become available over the past several years as a result of \textit{ab initio} simulations making use of density functional theory molecular dynamics \citep{2009PNAS..106.1324M,2009PhRvL.102k5701L,2011PhRvB..84w5109L,2013PhRvB..87q4105M}. The present work demonstrates that using \textit{ab initio} results for the H-He phase diagram, a differentiating non-adiabatic Jupiter comprised of hydrogen and helium surrounding a dense core of heavy elements explain Jupiter's evolutionary state at the solar age. 

To assess the viability of the evolution models, we formulate the problem in a Bayesian framework, using Jupiter's observed $\teff$, $\yatm$, and volumetric mean radius $\rvol$ to derive posterior probability distributions for the model parameters using a Markov chain Monte Carlo sampling algorithm. Most importantly, we  make a probabilistic determination of the superadiabatic temperature gradient to be expected in the deep interior, and simultaneously estimate the temperature correction that must be applied to the \cite{2011PhRvB..84w5109L} phase diagram to satisfy the \textit{Galileo} entry probe measurement of $Y_{\rm atm}$ \citep{2015MNRAS.447.3422N}. The present work thus extends the basic approach of \cite{2003Icar..164..228F}---using forward thermal evolution models to infer a most likely H-He phase diagram---with the power of a Bayesian parameter estimation method and the treatment of non-adiabatic $P-T$ profiles.

In \S\ref{s.models} we describe our modeling approach using Modules for Experiments in Stellar Astrophysics (\mesa; \citealt{2011ApJS..192....3P,2013ApJS..208....4P,2015ApJS..220...15P}), including our atmospheric boundary condition, treatment of helium phase separation, thermal transport, and other modifications that were necessary for our application.  We describe the three free parameters in our inhomogeneous, non-adiabatic models, namely the heavy-element core mass $\mc$, the double-diffusive superadiabaticity (or ``density ratio'') $\r$, and the phase diagram temperature offset $\dtdmx$. In \S3 we present results of evolutionary calculations, first validating our models for the case of homogeneous composition, then discussing in detail the late inhomogeneous, non-adiabatic evolution as a result of helium rain.  We then repeat these calculations, but treating the planet's equilibrium temperature $\teq$ as a fourth free parameter controlling the overall cooling time, mimicking the influence of a ``colder'' or ``warmer'' EOS than the adopted \cite{1995ApJS...99..713S} EOS. Marginalizing over this parameter allows us in an indirect sense to marginalize over the plausible H-He equations of state and thus obtain the most general estimates for the remaining three parameters. In \S4 we summarize and contextualize our findings.

\section{Planetary evolution models}\label{s.models}
Our evolution models are computed using Modules for Experiments in Stellar Astrophysics (\mesa; \citealt{2011ApJS..192....3P,2013ApJS..208....4P,2015ApJS..220...15P}).  The models are hydrostatic, nonrotating spheres  with envelopes consisting of binary mixtures of $^1$H and $^4$He surrounding dense inert cores of heavy elements. In the density-temperature regime relevant to giant planets with $M\lesssim\mj$, \mesa employs the \cite{1995ApJS...99..713S} equation of state, interpolated over hydrogen's molecular-metallic phase transition such that the density varies smoothly between the two phases (\scvhi). This EOS is advantageous for studies of helium phase separation because it provides the necessary state variables for arbitrary mixtures of hydrogen and helium, which is critical for solving the energy equation throughout the interior of the differentiating planet. This is one of our principal motivations for using \scvhi over more recent H-He EOSs obtained with \textit{ab initio} methods (e.g., \citealt{2013ApJ...774..148M}) in spite of the latter class comparing more favorably with shock experiments. Rosseland mean radiative opacities are taken from \cite{2008ApJS..174..504F} and a privately communicated 2011 update. Electron conduction opacities are based on \cite{2007ApJ...661.1094C}.

Jupiter is oblate as a result of its rapid rotation, while our present models are perfect spheres. A suitable mean planet radius with which to compare our model radii is the volumetric radius \citep{2007CeMDA..98..155S} 
\begin{equation}
	\rvol\equiv\req^{2/3}\rpol^{1/3}=69911\pm6\ {\rm km},\label{eq.rvol}
\end{equation}
where $\req$ and $\rpol$ are Jupiter's equatorial and polar radii at 1 bar; $\rvol$ is the radius of sphere enclosing the same volume as does Jupiter's 1-bar surface. Because the overall compactness of the planet is steeply sensitive to its heavy element mass, the high precision of this radius measurement translates into an extremely narrow range of allowed core masses. As an example, fitting the radius of our homogeneous, adiabatic Jupiter model---for which $\mc$ is the sole free parameter---to Jupiter's $\rvol$ at the solar age using MCMC produced $\mc=(25.33\pm0.03)\ \me$ (the quoted value corresponding to the median and the error to the 68\% confidence interval). Our models prefer large core masses because of the assumption that all heavy elements are relegated to a dense core; in reality, at least half of Jupiter's heavy element mass probably resides in the hydrogen-dominated mantle and envelope \citep{2004ApJ...609.1170S}. Thus in our findings (\S\ref{s.results}), $\mc$ should be interpreted as a total heavy element mass. Indeed, the fact that our adopted equation of state is limited to hydrogen and helium is the reason why we make no effort to calculate the oblateness and associated gravitational multipole moments $(J_2,\ J_4,\ \ldots)$ for our models.

The presence of heavy elements in the envelope and the action of rotation would both modify the hydrostatic structure and thus the total cooling time obtained for an evolutionary model. To assess the sensitivity to the heavy element distribution, we computed models with fixed total heavy element mass $\mz=\mc+\mzenv=28\ \me$ and varying combinations of the core mass $\mc$ and envelope heavy element content $\mzenv$, modelling the heavy elements in the H-He envelope simply by taking $Y\to Y+Z$ for the purposes of this diagnostic. We find that models with more of their heavy elements mixed throughout the envelope cool more quickly (e.g., \citealt{baraffe2008}), with a characteristic spread of 70 Myr in the time for models with no H-He phase separation to cool to Jupiter's volumetric radius. This is a relatively short time compared to the $\sim2-5\times10^8$ yr spread in cooling times obtained by varying the total heavy element content or atmospheric boundary condition, as discussed in \S3. The centrifugal support provided by rotation would also modify the radius evolution, but as there are a number of complexities associated with rotation (e.g., the likely differential rotation as a function of radius or latitude, and the details of the planetary figure as a result of even rigid rotation), we do not model rotation here. It should be noted that the effect of rotation is to some degree degenerate with the total amount and distribution of heavy elements, since the tendency for centrifugal acceleration to prefer larger radii can be offset by incorporating more heavy elements.

\subsection{Initial conditions}
Adiabatic, extended (1-bar radius $R=2\,\rj$) initial models of mass $M=\mj=1.89861\times10^{30}$ g were created using \mesa's \texttt{create\_initial\_model} capability described in \cite{2013ApJS..208....4P}. For each model, a fraction of the total mass was converted into a dense, nonevolving core of specified mass $\mc$ and density $\rhoc$. All models in the present work adopt a constant core density $\rhoc=15$ g cm$^{-3}$ for simplicity. The evolution is insensitive to the particular choice of core density; the total radius $R$ is set by $\mc$ nearly independent of $\rhoc$.

\subsection{Hydrogen-helium phase separation}
We couple our evolutionary calculations to the H-He phase diagram of \cite{2011PhRvB..84w5109L}, whose calculations include the simplifying assumption of ideal entropy of mixing between the two species.  The group of \cite{2013PhRvB..87q4105M} instead performed direct thermodynamic integrations, thus including nonideal contributions to the entropy of mixing, but at the expense of subtantially more sparse sampling in helium number fraction $\xhe$. The results of the two groups are in reasonable agreement, diverging most noticeably at temperatures $\lesssim 3000$ K, which on a present-day Jupiter (or Saturn) adiabat is well outside the transition from molecular to metallic hydrogen predicted by either group. Although it appears from their Figure 1 that the phase diagram of \cite{2013PhRvB..87q4105M} predicts that Jupiter has not yet cooled into the immiscibility region of $(P,\ T)$ space for a protosolar mixture, the favored model of \cite{hubbard2016} has a cooler deep interior such that it has already undergone H-He phase separation according to both phase diagrams. These cooler deep layers are a result that \cite{2008ApJ...688L..45M} attributed to the important inclusion of the nonideal entropy of mixing between H and He.

Apart from the major differences in the two modern phase diagrams' behavior at $T\lesssim3000$ K, $P\lesssim1$ Mbar, their overall shapes at the warmer conditions relevant to Jupiter and Saturn are in rough agreement, with the major effect being an temperature offset (up to $1000$ K) between the two.  It is possible that this discrepancy stems from including versus excluding the nonideal entropy of mixing between the two components. It was demonstrated by \cite{2015MNRAS.447.3422N} that applying the raw phase diagram of \cite{2011PhRvB..84w5109L} to Jupiter resulted in too large a loss of helium from the molecular envelope, motivating a global downward offset in the demixing temperature such that the onset of phase separation takes place later in the planet's history.  It is reassuring that an offset motivated by Jupiter's observed atmospheric helium content brings the two phase diagrams into closer agreement.  In this paper, we also allow for a global temperature offset $\dtdmx$ to this phase diagram and estimate its value from available data; we confirm the result of \cite{2015MNRAS.447.3422N} that for models computed with \scvhi, the necessary downward offset is of order 200 K. Shifts to the published phase diagram in pressure are also conceivable, but as the phase diagrams of \cite{2011PhRvB..84w5109L} and \cite{2013PhRvB..87q4105M} agree fairly closely on the minimum \textit{pressure} for phase separation at temperatures typical of Jupiter and Saturn's molecular-metallic transitions, the temperature offset appears to be the most important correction.

\begin{figure}[ht]
	\begin{center}
		\includegraphics[width=\columnwidth]{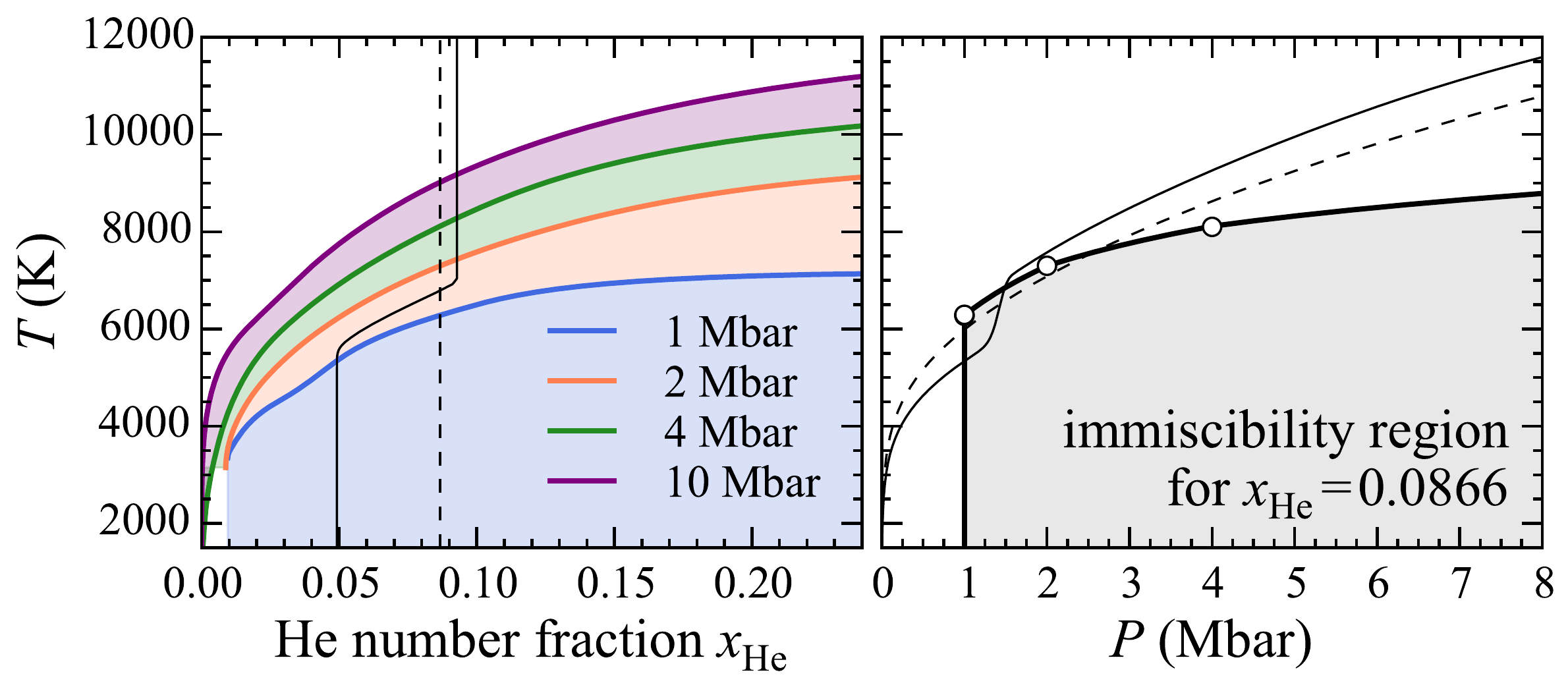}
		\caption{\label{fig.phase_diagram}
		H-He phase diagram of \cite{2011PhRvB..84w5109L}, illustrating immiscibility regions as a function of helium number fraction $\xhe$ for four pressures (left panel), and as a function of pressure for the protosolar mixture (right panel).  Immiscibility regions (shaded regions bounded by thick curves) are precluded in terms of equilibrium thermodynamics, i.e., by the criterion that the Gibbs free energy be stable with respect to perturbations in the helium concentration.  The vertical dashed line in the left panel designates the protosolar mixture, $\xhe=0.0866$ ($Y=0.275$). Open circles in the right panel are raw data and the curve is a linear interpolation in $\log P$.  A solar-age profile of a representative differentiated Jupiter ($\mc=30\,\me$, $\r=0.25$, $\dtdmx=0$) is shown in the thin black solid curves.}
	\end{center}
\end{figure}

The phase diagram of \cite{2011PhRvB..84w5109L} adopted for the models in this paper is illustrated in Figure~\ref{fig.phase_diagram}, wherein phase curves are shown in $(\xhe,\ T)$ space (left panel) and in $(P,\ T)$ space (right panel). The data are finely sampled in $\xhe$ but more coarsely in $P$, where data are available at $P=(1,\ 2,\ 4,\ 10,\ 24)$ Mbar. Our model assumes no phase separation at pressures lower than $1\ {\rm Mbar}$ where no data are available. This assumption is reasonable given the near-vertical $(P,\ T)$ phase curves found by both \cite{2011PhRvB..84w5109L} and \cite{2013PhRvB..87q4105M} for the relevant temperatures $4\ {\rm kK} < T < 6\ {\rm kK}$, although the latter phase diagram situates the boundary at slightly lower pressure $P\approx0.8\ {\rm Mbar}$. To compute the equilibrium abundances for zones at arbitrary $T$ and $P$, we interpolate in the tables linearly in $\xhe$ and $\log P$ to obtain $\tdmx$, and numerically solve the equation \begin{equation}
	\tdmx(\xhe,\ P)+\dtdmx-T=0\label{eq.tdmx}
\end{equation}
for the equilibrium helium concentration $\xhe$. 

Differentiation begins once the planetary $P-T$ profile cools into the immiscibility region for the protosolar mixture, at which time one or more grid points are supersaturated in helium. For this and each subsequent timestep, we assume the thermodynamic equilibrium distribution of helium throughout the interior, which amounts to the assumption that all helium excess is delivered efficiently by gravitational settling to lower depths such that it can exist in chemical equilibrium with its surroundings. Here ``efficient'' delivery means that the local supersaturation can be reduced to zero faster than the other relevant timescales, namely the planet's thermal timescale, the large-scale convective circulation time, and the simulation timestep. 

The thermal timescale $\tau_{\rm th}=E_{\rm gr}/L$, with $E_{\rm gr}$ denoting the total binding energy, is roughly $10^9\ {\rm yr}$. Timesteps in our simulations are typically $10^6$ to $10^7$ yr. The convective circulation time presents the strongest condition: it can be estimated with mixing length theory as \citep{2004jpsm.book...35G} $\tau_{\rm MLT}\sim10^8\ {\rm s}$, or 3 yr. \cite{ss77b} derived the minimum size a He-rich droplet would need to attain in order for its terminal speed to exceed the average speed of convective motions of the ambient fluid. Balancing the terminal speed from a turbulent drag law with an average convective speed estimated from mixing length theory, those authors obtained a minimum droplet length scale on the order of 1 cm. Repeating their calculation, we balance the (downward) buoyancy force with a drag force through a convective plume:
\begin{equation}
	\Delta\rho d^3 g\approx C\rho d^2v^2,\label{eq.force_balance_turbulent}
\end{equation}
where $\Delta\rho$ is the density excess compared to the surrounding H-dominated fluid, $d$ is the length scale of the droplet, $C$ is a nondimensional drag coefficient, and $v$ is the relative speed between the droplet and the medium. A minimum droplet size $d_{\rm min}$ corresponds to a droplet having $v=v_{\rm MLT}$ so that
\begin{equation}
	d_{\rm min} \approx \frac{C\rho v_{\rm MLT}^2}{\Delta\rho g},
\end{equation}
where $v_{\rm MLT}\approx 10\ {\rm cm\ s}^{-1}$, $\Delta\rho\approx\rho$, and $g\approx3\times10^3\ {\rm cm\ s}^2$.  If as in \cite{ss77b} we assume turbulent flow over the scale of a droplet (${\rm Re}\equiv vd/\nu\gtrsim10^3$), then $C=0.5$ \citep{1959flme.book.....L} and we arrive at a smaller minimum size $d_{\rm min}\approx10^{-2}$ cm; the discrepancy may be the result of a dimensional error in the earlier calculation. However, given a typical kinematic viscosity $\nu=4\times10^{-3}\ {\rm cm}^2\ {\rm s}^{-1}$ \citep{2012ApJS..202....5F}, the Reynolds number ${\rm Re}\equiv ud/\nu\approx25$ so that a turbulent drag force is not appropriate.  If instead we balance buoyancy with a Stokes drag appropriate for low Reynolds number, then \citep{1959flme.book.....L}
\begin{equation}
	\Delta\rho d_{\rm min}^3g\approx 6\pi d_{\rm min}\rho v_{\rm MLT}\nu
\end{equation}
and we obtain
\begin{equation}
	d_{\rm min}=\left(\frac{6\pi\nu v_{\rm MLT}}{g}\right)^{1/2}\approx 10^{-2}\ {\rm cm},
\end{equation}
incidentally the same estimate as in the turbulent case.

Again following \cite{ss77b}, this droplet size can be translated into a droplet formation timescale if we suppose that the droplets grow by microscopic diffusion of He nuclei into the He-rich pockets. Assuming the microscopic diffusivity $D_{\rm He}$ to be of order the He-He self-diffusion coefficient obtained in the recent \textit{ab initio} simulations by \cite{2012ApJS..202....5F}, the timescale to form a sinkable droplet is
\begin{equation}\label{eq.tau_droplet}
	\tau_{\rm sedimentation}=\frac{d_{\rm min}^2}{D_{\rm He}}=\frac{(10^{-2}\ {\rm cm})^2}{10^{-3}\ {\rm cm}^2\ {\rm s}^{-1}}\approx 10^{-1}\ {\rm s}.
\end{equation}
The strong hierarchy in timescales
\begin{equation}
	\tau_{\rm sedimentation} \ll \tau_{\rm MLT} \ll \tau_{\rm th}
\end{equation}
implies that the transport of excess helium toward the center of the planet is probably efficient in spite of convection (or double-diffusive convection, which mixes material over much longer timescales), so assuming the equilibrium mixture throughout the planet at each timestep is an adequate starting point for evolutionary models.

For all timesteps in which a cell has been cooled below its critical temperature, we lower the abundance of the outermost supersaturated grid point (i.e., the first grid point with $P>1\ {\rm Mbar}$) to its equilibrium value.  We apply this same equilibrium abundance throughout the molecular envelope, reflecting the fact that outside of regions where phase separation is taking place, the species are being rapidly mixed by convection. Since a single timestep corresponds to millions of large-scale convection cycles, the entire molecular envelope acts as a reservoir of helium for the phase separation and rainout taking place near the molecular-metallic transition. The molecular envelope thus depletes uniformly in a given step.

Iterating inward over grid points, we enforce the local equilibrium abundance (a monotonically increasing function of depth) in each cell, and propose that same abundance as the tentative (He-enriched) mixture to be applied to the remainder of the metallic interior. Eventually, a grid point is reached whose equilibrium abundance is greater than or equal to its proposed abundance. At this point the inward iteration ceases and the homogeneous, He-enriched interior is stable to phase separation; all that remains is to enforce conservation of helium overall. The proposed profile always has a helium deficit, which is resolved by the constraint that all helium nuclei that have rained out from above are mixed into the homogeneous interior; we raise the helium abundance of the homogeneous interior accordingly. Since this adjustment typically leaves the uppermost layers in the homogeneous interior marginally supersaturated, we repeat the iteration over all grid points as many times as necessary to achieve the equilibrium profile. In practice this takes one or two more iterations.

\subsection{Modes of heat transport}\label{s.semiconvection_model}
The pioneering work of \citet{hubbard68, hubbard69, hubbard70} made the case for Jupiter's envelope as a convective fluid of primarily hydrogen and helium. The short mean free path of photons and electrons in the interior imply a small thermal conductivity, with the result that convection, rather than radiation or conduction, carries nearly all the intrinsic flux. The temperature gradient is thus marginally superadiabatic; estimates from mixing length theory yield $\grad-\grada\lesssim10^{-6}$ throughout the envelope and $\grad-\grada\lesssim10^{-9}$ within hydrogen's molecular-metallic phase transition at megabar pressures. Our homogeneous Jupiter models indicate that convection maintains these small superadiabaticities to within an order of magnitude over the history of the planet with the exception of an early, brief radiative window similar to that described by \cite{1995ApJ...450..463G}. As demonstrated by \cite{2004jpsm.book...35G}, the inclusion of alkali metals enhances the opacity enough to ensure convection at all depths within a present-day Jupiter. Our models, which include modern opacities and self-consistently allow for radiative transport wherever $\gradr<\grada$, confirm this result: the intermediate radiative window is only mildly subadiabatic ($\grad-\grada\sim-10^{-1}$) and vanishes before $t\sim2\times10^8$ yr (see also \citealt{fortney2011}), and has an insignificant effect on the overall cooling time.  Thus for the homogeneous phases of the evolution, our models may be directly compared to models constructed assuming $\grad=\grada$ always.

In the more general case allowing for stratification in the mean molecular weight $\mu$, the \cite{1958ApJ...128..348S} criterion for dynamical convection
\begin{equation}
	\grad > \grada \label{eq.schw}
\end{equation}
must be replaced by the \cite{1947ApJ...105..305L} criterion
\begin{equation}
	\grad > \grada+\frac\varphi\delta\gradmu,\label{eq.ledoux}
\end{equation}
where
\begin{equation}
	\gradmu\equiv\frac{d\ln\mu}{d\ln P}\label{eq.gradmu_def}
\end{equation}
is the slope of the mean molecular weight $\mu$ along the planetary profile, and 
$\varphi$ and $\delta$ are two thermodynamic derivatives defined by
\begin{equation}
	\varphi=\left(\frac{\partial\ln\rho}{\partial\ln\mu}\right)_{P, T}, \quad
	\delta=-\left(\frac{\partial\ln\rho}{\partial\ln T}\right)_{P,\mu}.
\end{equation}
(Details on the novel calculation of $(\varphi/\delta)\gradmu$ in \mesa are given in \citealt{2013ApJS..208....4P} \S3.3). If mean molecular weight increases toward the planet's center, as is the case for a differentiated planet, then $\gradmu>0$ and the regime
\begin{equation}
	\grada < \grad < \grada+\B \label{eq.semiconvection}
\end{equation}
corresponds to the situation wherein a superadiabatic temperature profile is dynamically stabilized by the chemical stratification. The mixing that ensues in this regime is termed ``semiconvection'' in the stellar context \citep{1958ApJ...128..348S,1974ApJ...190..101S} and ``double-diffusive convection'' in the hydrodynamic context \citep{1974AnRFM...6...37T}. In this case the \bv frequency $N$ of the fluid is real-valued, admitting gravity waves. These modes are in general overstable if fluid parcels can exchange a significant fraction of their heat with the environment over an oscillation period. The linear stability analysis of \cite{1966PASJ...18..374K} demonstrated that in the stellar case, where radiative diffusion is efficient, the thermal diffusion timescale tends to be short compared to buoyant oscillation periods $N^{-1/2}$ so that the criterion for convective instability reduces to Equation~\ref{eq.schw} and marginally superadiabatic temperature gradients can be sustained by convection. \cite{1979MNRAS.187..129S} came to the same conclusion, arguing that the otherwise weak vertical heat transport provided by these overstable oscillations is mitigated by the occasional breaking of waves \citep{2011ApJ...731...66R}, redistributing solute such that $\grad=\grada$ to a good approximation. 

Metallic hydrogen environments in cool giant planets differ from the stellar case for two reasons: (i) the formation, rainout, and deeper redissolution of helium droplets tends to enforce a persistent, stabilizing composition gradient, and (ii) since the (conduction-limited) diffusion of heat is quite inefficient, overstable gravity waves have relatively slow growth rates so that wave-breaking events are rare and the fluid is only weakly turbulent; vertical heat transport is thus enhanced relative to the purely diffusive case but is still much weaker than in the case of overturning convection. The resulting temperature gradient is substantially superadiabatic, possibly closer to the Ledoux limit $\grad=\grada+(\varphi/\delta)\gradmu$ \citep{1979MNRAS.187..129S,2012ApJ...750...61M}. 

\subsection{A parametric model for double-diffusive convection}\label{s.ddc_param}
\begin{figure}[ht]
	\begin{center}
		\includegraphics[width=0.75\columnwidth]{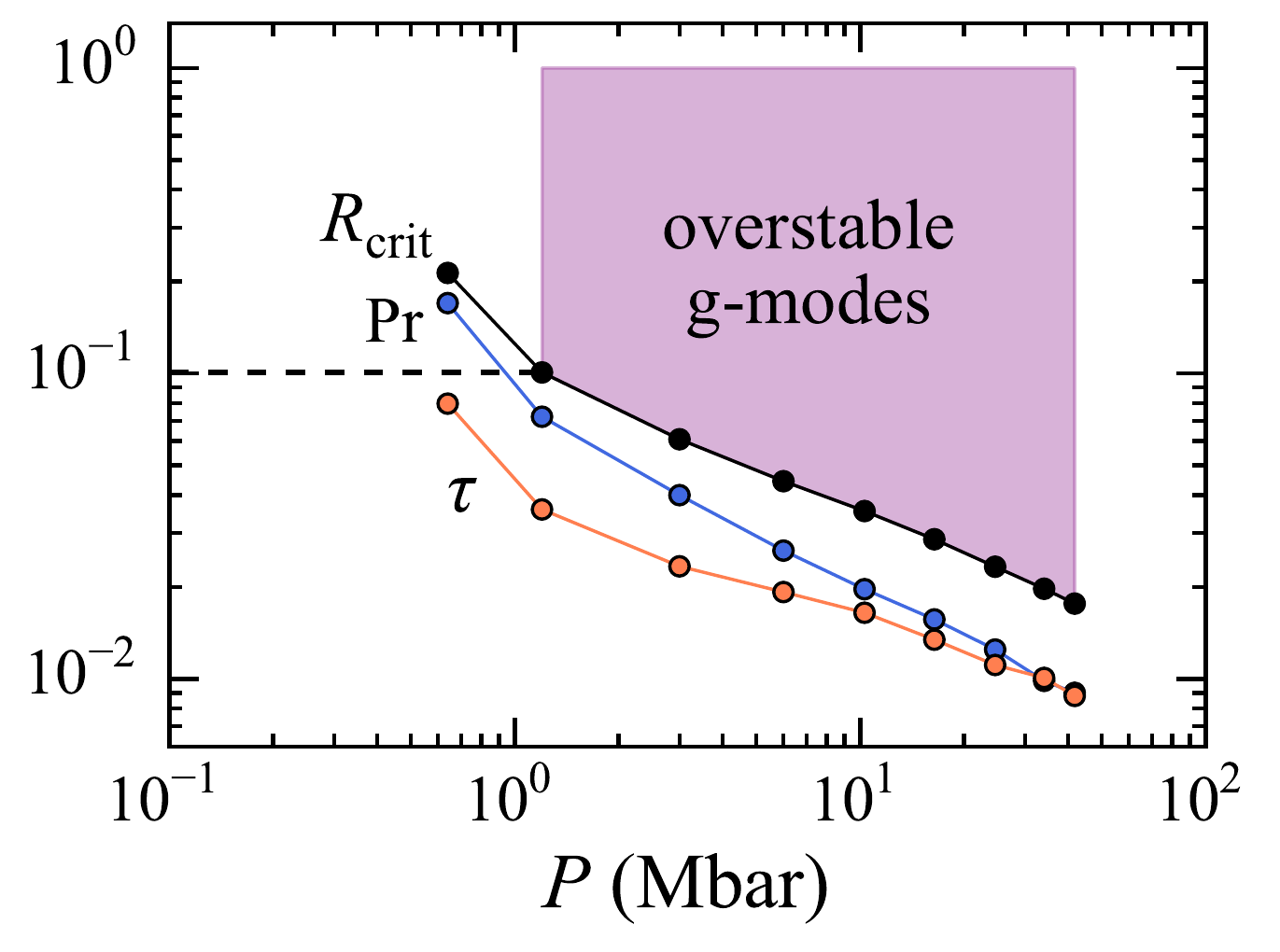}
		\caption{\label{fig.pr_tau_rcrit} Estimates of the dimensionless quantities $\pr$ and $\tau$ (Equation~\ref{eq.pr_tau}), as well as the critical density ratio $\rcrit$ for overstable double-diffusive convection (Equation~\ref{eq.rcrit}).  Quantities are derived from the \textit{ab initio} transport properties of \cite{2012ApJS..202....5F} for the metallic hydrogen part of Jupiter's interior. The shaded region is the intersection of $\rcrit<\r<1$ and $P>1\ {\rm Mbar}$, within which the stable stratification from helium rain admits growing-amplitude gravity waves and thus some degree of double-diffusive convection.}
	\end{center}
\end{figure}

We assume the temperature gradient $\grad$ to be adiabatic unless there exists a stabilizing composition gradient $\gradmu>0$, in which case the temperature gradient is steeper than the adiabat by an amount proportional to $\gradmu$, i.e.,
\begin{equation}\label{eq.r0}
	\grad=\grada+\r\frac\varphi\delta\gradmu\ \Longleftrightarrow\  \r=\frac{\grad-\grada}{(\varphi/\delta)\gradmu}.
\end{equation}
Equation~\ref{eq.r0} defines the density ratio $\r$, which describes the relative (and competing) contributions that the temperature and composition stratifications make to the overall density stratification. We take $\r$ as a free parameter that we seek to estimate. Although the commonly cited criterion of Equation~\ref{eq.semiconvection} is a necessary condition for semiconvection, it is not sufficient; the linear instability demands the somewhat more strict criterion \citep{1964Tell...16..389W,1966PASJ...18..374K}
\begin{align}
	1 > \r > \frac{\pr+\tau}{\pr+1}\equiv \rcrit.\label{eq.rcrit}
\end{align}
Thus for the semiconvective instability to grow, the temperature gradient must be superadiabatic by a nonvanishing amount determined by the Prandtl number $\pr$ and diffusivity ratio $\tau$ defined by
\begin{equation}
	{\rm Pr}=\frac{\nu}{\kappa_T},\qquad \tau=\frac{\kappa_\mu}{\kappa_T},\label{eq.pr_tau}
\end{equation}
where $\nu$ is the fluid's kinematic viscosity, $\kappa_T$ is its thermal diffusivity, and $\kappa_\mu$ is the diffusivity of solute, in this case the diffusivity of helium atoms in a mixture which is predominantly metallic hydrogen. In Figure~\ref{fig.pr_tau_rcrit} we show values of $\pr$, $\tau$, and $\rcrit$, derived from the \textit{ab initio} transport properties obtained by \cite{2012ApJS..202....5F} for a Jupiter adiabat.  Here the calculation of $\tau$ (and thus $\rcrit$) assumes an effective composition diffusivity $\kappa_\mu$ that is of order the He-He self-diffusion coefficient reported in that paper. The values of $\rcrit$ indicate that overstable modes can grow at 1 Mbar and deeper for density ratios $\r>10^{-1}$.

\subsection{Energetics of evolving compositions}
Enriching or depleting a Lagrangian fluid element in helium generally modifies its internal energy per gram $u$ and does work modifying its density $\rho$ following the fundamental thermodynamic relation
\begin{equation}\label{eq.thermodynamic}
	T\frac{ds}{dt}=\frac{du}{dt}+P\frac{d(1/\rho)}{dt},
\end{equation}
where $s$ is the entropy per gram and $d/dt$ denotes a time derivative.  This change in heat content is commensurate with the energy gained or lost by the fluid element via photons:
\begin{subequations}\label{eq.energy}
\begin{align}
	\frac{dL}{dm}&=-T\frac{ds}{dt} \label{eq.tds} \\
	&=-\frac{du}{dt}-P\frac{d(1/\rho)}{dt}. \label{eq.pdv}
\end{align}
\end{subequations}
Here $L$ is the local luminosity and $m$ is the Lagrangian coordinate; other energy sources and sinks (nuclear fusion or fission, tidal dissipation, neutrino cooling) are negligible for our application. While the two forms of the energy equation Eqs.~\ref{eq.tds} and \ref{eq.pdv} are fundamental, \mesa's solvers do not work with the entropy directly, and also do not work with $(u,\ \rho)$ by default (although the latter option exists). In either of the two standard thermodynamic bases $(\rho,\ T)$ or $(P,\ T)$, these two variables (along with $m$, $r$, and $X_i$) are solved for simultaneously and then $s$ or $u$ are computed from the EOS post-hoc. Because $s$ or $u$ are not solved for directly, their finite differences over time are subject to numerical noise, and direct finite differences of either form \ref{eq.tds} or \ref{eq.pdv} thus yield noisy luminosity profiles. The approach \mesa takes by default is to instead recast Equation~\ref{eq.energy} into time derivatives of the quantities comprising the adopted thermodynamic basis.  Since the phase diagram of \cite{2011PhRvB..84w5109L} provides the equilibrium helium abundance over $(P,\ T)$ space, and these are also the independent variables in the \cite{1995ApJS...99..713S} EOS, we choose to adopt $(\ln P,\ \ln T)$ as the thermodynamic basis for our \mesa calculations. In this basis the energy equation takes the form \citep{2013ApJS..208....4P}
\begin{equation}\label{eq.energy_PT}
	\left.\frac{dL}{dm}\right|_{X_i}=-c_PT\left(\frac{d\ln T}{dt}-\grada\frac{d\ln P}{dt}\right),
\end{equation}
where we have assumed radiation pressure is negligible as is appropriate for $T\lesssim 10^4$ K.

The standard transformation of Equation~\ref{eq.energy} into Equation~\ref{eq.energy_PT} (e.g., \citealt{1990sse..book.....K}) ignores the fact that entropy and internal energy depend not only on $P$ and $T$ but also on the composition vector $X_i$, and hence Equation~\ref{eq.energy_PT} is only accurate at fixed composition.  This poses no substantial problem for energy conservation in stellar models, where large abundance changes typically only happen as a result of fusion, in which case nuclear energy generation overwhelms the $T({ds}/{dt})$ term in the energy equation; one important exception is the accretion of material with a composition different from the stellar surface. For our application, it is necessary to add to Equation~\ref{eq.energy_PT} the component of $dL/dm$ that arises from composition changes at fixed $P$ and $T$:
\begin{equation}\label{eq.eps_grav_comp_full}
	\left.\frac{dL}{dm}\right|_{P,\ T}=-\frac{\partial u}{\partial X_i}\frac{dX_i}{dt}-P\frac{\partial(1/\rho)}{\partial X_i}\frac{dX_i}{dt}.
\end{equation} 
Here the repeated indices denote summation over species $i=1,\ \ldots,\ N-1$ where $N$ is the total number of species in the model. (Since all $N$ mass fractions sum to unity, only $N-1$ mass fractions are independent.)  All models in this work assume a two-component mixture of $^1$H and $^4$He, so that Equation~\ref{eq.eps_grav_comp_full} reduces to just
\begin{equation}
	\label{eq.eps_grav_comp}
	\left.\frac{dL}{dm}\right|_{P,\ T}=-\frac{\partial u}{\partial Y}\frac{dY}{dt}-P\frac{\partial(1/\rho)}{\partial Y}\frac{dY}{dt}.
\end{equation}
In practice we calculate this term for each cell as
\begin{align}
	\label{eq.eps_grav_comp_finite}
	\left.\frac{dL}{dm}\right|_{P, T}=&-\frac{u(P, T, Y_0)-u(P,T,Y_1)}{\Delta t} \nonumber \\
	&-P\left(\frac{\rho^{-1}(P, T, Y_0)-\rho^{-1}(P, T, Y_1)}{\Delta t}\right)
\end{align}
where $Y_0$ and $Y_1$ denote the helium mass fractions before and after our helium redistribution step, and $\Delta t$ denotes the (finite) timestep. Obtaining $u(P, T, Y_1)$ and $\rho^{-1}(P, T, Y_1)$ requires one additional call to the EOS module per zone per timestep.

\subsection{Model atmospheres}\label{s.atmospheres}
The overall cooling time of an evolutionary giant planet model depends strongly on the boundary condition applied at the atmosphere, since that condition determines how rapidly the planet can radiate.  We apply the self-consistent model atmospheres of \cite{fortney2011} as fit analytically by \cite{lc13}, which provide the temperature at the 10 bar level $T_{10}$ as a function of surface gravity $g$ and intrinsic temperature $T_{\rm int}$. The planet's effective temperature $T_{\rm eff}$ in a given timestep is given by
\begin{equation}
	T_{\rm eff}^4 = T_{\rm int}^4 + T_{\rm eq}^4,
\end{equation}
where the equilibrium temperature $T_{\rm eq}$ describes the orbit-averaged temperature of a body radiating as much energy as it absorbs from the Sun following
\begin{equation}
	\sigma T_{\rm eq}^4 = \frac{(1-A)L}{16\pi a^2},
\end{equation}
with $L$ the instantaneous luminosity of the star and $a$ the planet's orbital semimajor axis. Effective temperatures for our models are calculated assuming the value for $A$ determined by Voyager measurements in the infrared \citep{1981Sci...212..192H,1991JGR....9618921P}. Although the planet's albedo is certainly a function of time as a result of changing atmospheric dynamics and chemistry \citep{1973SSRv...14..599H}, assuming the present-day value of the albedo throughout the evolution is an acceptable (and at this stage necessary) approximation. 

While Voyager 1 measured a Bond albedo $A=0.343\pm0.032$, corresponding to $T_{\rm eq}=109.9\pm1.3$ K, \cite{1991JGR....9618921P} noted that the value of $A$ determined using Voyager 2 radiometry is larger than that determined by Voyager 1 by roughly 12\%.  As those authors suggested, if this discrepancy is the result of unidentified systematic error such that the true value is $6\pm6\%$ larger than the Voyager 1 measurement, then a revised estimate for Jupiter's Bond albedo is $A=0.366\pm0.035$. This corresponds to a modestly smaller equilibrium temperature $T_{\rm eq} = 108.5\pm1.4$ K.  In the present work we adopt the median equilibrium temperature $T_{\rm eq}=109.0$ K of the two Voyager determinations as ground truth. In \S\ref{s.mcmc} we also show our full calculations repeated with $T_{\rm eq}$ as a free parameter to address the broad modeling uncertainties---principally those associated with the EOS---that contribute to the uncertain overall cooling time for a contracting giant planet.

\begin{figure}[ht]
	\begin{center}
		\includegraphics[width=\columnwidth]{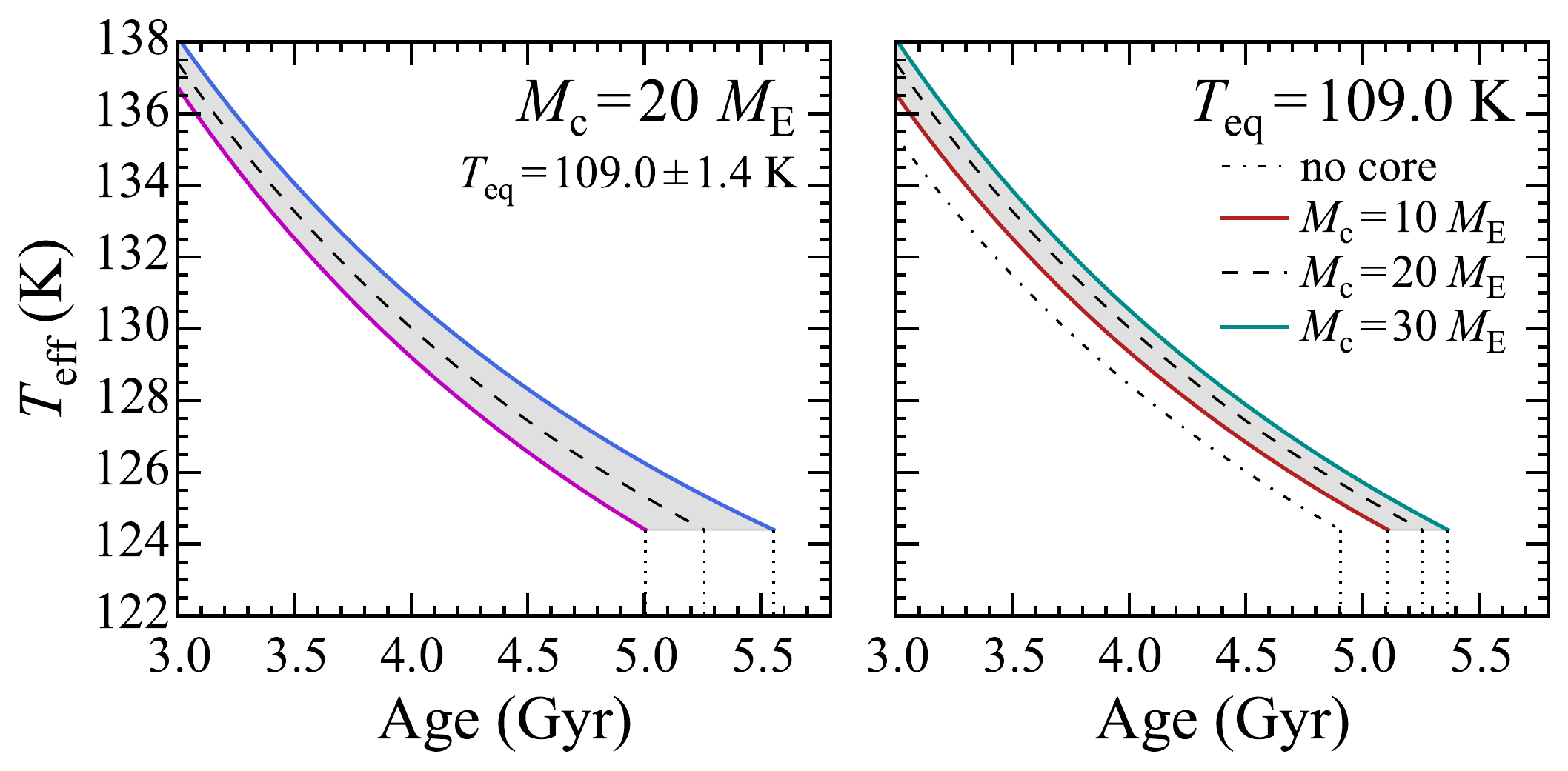}\label{fig.homogeneous_tracks}
		\caption{\label{fig.homog_tracks}Sensitivity of the cooling time to the surface boundary condition (left panel) and heavy-element mass (right panel) for homogeneous, adiabatic Jupiter models. The dashed curve corresponds to $T_{\rm eq}=109.0$ K, and the blue (magenta) curve corresponds to plus (minus) $\sigma_{\teq}=1.4$ K.}
	\end{center}
\end{figure}

\section{Results of Evolutionary Calculations}\label{s.results}
\begin{figure}[ht]
	\begin{center}
		\includegraphics[width=\columnwidth]{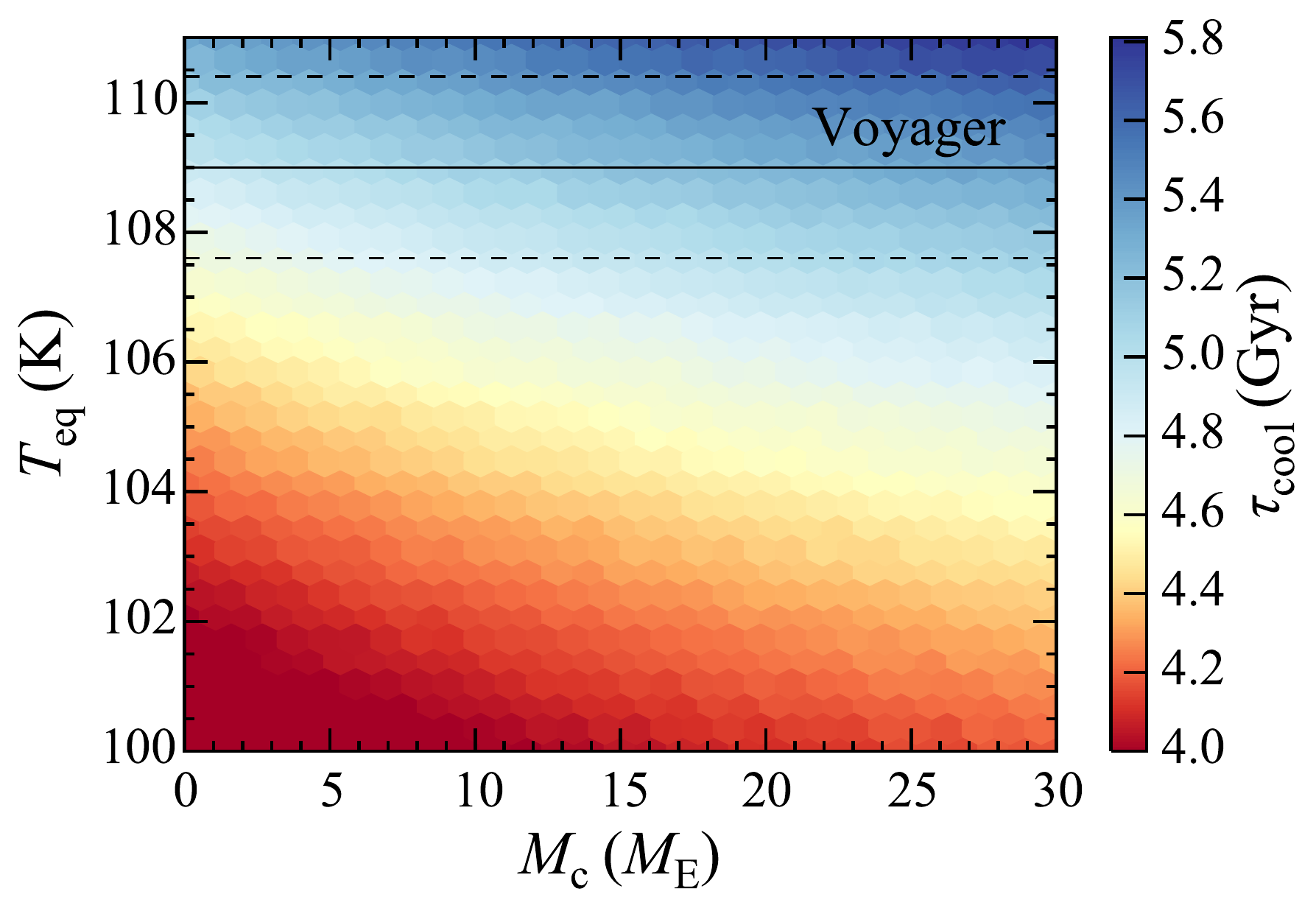}
		\caption{\label{fig.homog_cooling_times}Time for homogeneous $1.0\ \mj$ models to cool to Jupiter's $\teff$ as a function of their heavy-element mass $\mc$ and equilibrium temperature $\teq$. The horizontal lines designate the Voyager measurement of $\teq$ (see \S\ref{s.atmospheres}). The color scale is piecewise linear such that yellow corresponds to the solar age 4.56 Gyr.}
	\end{center}
\end{figure}

We first validate our general modeling approach and implementation of the model atmospheres by computing homogeneous, adiabatic evolutionary sequences.  Figure~\ref{fig.homog_tracks} shows evolution in the age-$\teff$ plane for models with core masses between 0 and $30\ \me$ and a range of assumed equilibrium temperatures reprensenting the uncertainty in the Voyager determination of Jupiter's Bond albedo. Models with higher equilibrium temperatures generally take longer to cool because they absorb more stellar flux, and models with greater heavy element content generally cool faster because they are more compact. Figure~\ref{fig.homog_cooling_times} provides a summary of cooling times attainable by homogeneous models across $\mc-\teq$ parameter space, and demonstrates that over the cooling time depends more steeply on the atmospheric boundary condition than on the assumed heavy element mass. The total cooling times agree closely with published results also using the \scvhi EOS \citep{2003Icar..164..228F,2004ApJ...609.1170S,fortney2011}.

The \scvhi EOS generally leads to slow cooling for the homogeneous models. These were evolved to Jupiter's observed $\teff=124.4$ K, a temperature notably not reached within the age of the solar system (4.56 Gyr) for any of the models, even those with vanishing heavy element mass.  It is clear that whatever superadiabaticity arises from helium rain in the inhomogeneous case must act to \textit{accelerate} the planet's cooling in spite of any additional luminosity associated with differentiation. This is the first indication that inhomogeneous models computed with the \scvhi equation of state will tend to favor fairly weak heat transport in the helium gradient region such that the temperature distribution is strongly superadiabatic and a thermal boundary layer is established. As described by \cite{ss77b}, in this case the molecular envelope can cool rapidly while the cooling of the deeper interior is stalled or even reversed. Indeed, sections~\ref{s.inhomog} and \ref{s.mcmc} below demonstrate that best-fitting models have a steep enough temperature gradient in the stable region that the cooling of the molecular envelope is accelerated while the helium-enriched metallic interior is heated over time.

Some more recent equations of state based on \textit{ab initio} methods (e.g., \citealt{2013ApJ...774..148M}) yield adiabats which are colder at Mbar pressures and Jupiter- and Saturn-like entropies than the adiabats obtained from the semianalytic model of \cite{1995ApJS...99..713S}. These ``colder'' equations of state compare more favorably with shock experiment, and generally predict shorter cooling times for homogeneous Jupiters because their adiabats have less total internal energy for a given global entropy. With such an EOS, if inhomogeneous evolution is to help with addressing Jupiter's luminosity constraint, then any double-diffusive convection in the deep interior must act to \textit{lengthen} the cooling relative to the homogeneous case. Thus for a ``colder'' equation of state, more modest superadiabaticities should be expected such that the differentiation luminosity overwhelms the accelerated cooling of the envelope due to the double-diffusive bottleneck at 1-2 Mbar. This situation is closer to the luminosity problem for Saturn, where the drastic underluminosity of homogeneous models is robust with respect to the assumed equation of state. In \S\ref{s.inhomog} we illustrate the central role that the efficiency of heat transport by ODDC plays in determining the thermal evolution. In \S\ref{s.mcmc} we retrieve strong superadiabaticities for our nominal \scvhi case. Then to address the systematic modelling uncertainty associated with the H-He equation of state, we repeat our calculations with $\teq$ taken as a fourth free parameter to modulate the overall cooling time, with high (low) $\teq$ mimicking the effect of a warmer (colder) equation of state.

\subsection{Inhomogeneous evolution}\label{s.inhomog}
\begin{figure*}[ht]
	\begin{center}
		\includegraphics[width=\textwidth]{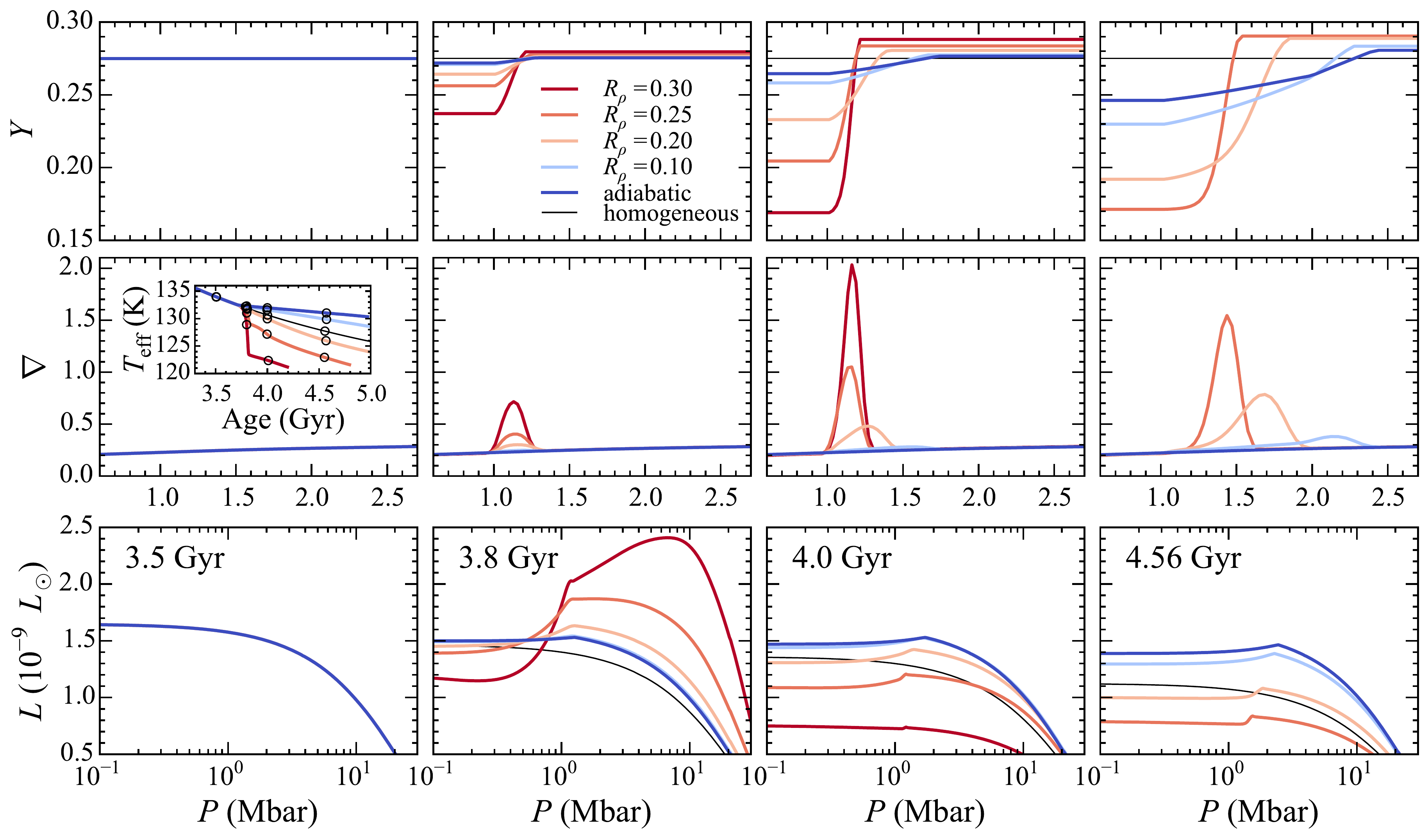}
		\caption{Interior profiles for differentiating Jupiter models. Shown are the helium mass fraction $Y$ (top row), the temperature gradient $\grad\equiv d\ln T/d\ln P$ (middle row; see Equation~\ref{eq.r0}) and local luminosity $L$ (bottom row) as functions of pressure for $1.0\ \mj$ models with $\mc=30\ \me$ and $\dtdmx=0$ K, for four different values of the fractional superadiabaticity $\r$: 0.20 (light orange), 0.25 (dark orange), 0.30 (red), and the adiabatic case $\r=0$ (blue). Thin black curves show a model with no phase separation for reference. The four columns correspond to four points in model age as labeled in text in the lower panels, and as indicated by the open circles in the inset in the center left panel. To emphasize detail, the first two rows show $Y$ and $\grad$ over a different pressure scale than the luminosity panels.}
		\label{fig.he4_grad_lum_profiles}
	\end{center}
\end{figure*}

Interior profiles for Jupiter models undergoing helium rain are illustrated in Figure~\ref{fig.he4_grad_lum_profiles} for four different values of $\r$, including the adiabatic case $\r=0$. (In the case of evolving composition profiles, $\grada=\grada(P,\ T,\ X_i)$ where the $X_i$ are now functions of depth. We refer to the case where $\grad=\grada$ everywhere as adiabatic, although the profiles are emphatically not isentropic.) For this figure and all others in this section, unless otherwise indicated, the two remaining free parameters are arbitrarily chosen as $\mc=30\ \me$ and $\dtdmx=0~{\rm K}$ for illustrative purposes. The first column shows profiles at 3.5 Gyr, at which time helium rain has not yet commenced and the models are thus still in an identical state.  The remaining three panels show profiles shortly after the onset of helium rain (3.8 Gyr), then at a more typical time after helium rain onset (4.0 Gyr), and finally at the solar age (4.56 Gyr).  The inset in the center left panel plots $\teff$ as a function of age for the four models. 

Evolution in the adiabatic case $\r=0$ is simplest because there is minimal feedback between the evolving composition and temperature profiles, and thus the shape of $L(m)$ can be understood by inspecting just the composition term of $dL/dm$ (Equation~\ref{eq.eps_grav_comp}); the thermal term retains essentially the same smooth profile as before helium rain sets in.  The layers with decreasing helium abundance---from the planetary surface down to the lower boundary of the helium gradient region---have an \textit{increasing} internal energy $u$ and specific volume $1/\rho$, and thus $dL/dm<0$ there.  Similarly, layers deeper than the bottom of the gradient region have a uniformly increasing helium abundance and thus $dL/dm>0$ there.  Hence, a global maximum in the luminosity is attained at the base of the gradient region. The adiabatic model also exhibits the most extended helium gradient region among the models considered (spanning 1 to 2.4 Mbar at the solar age), simply because a shallower $T(P)$ profile intersects the immiscibility gap over a broader range in pressure. Likewise, larger values of $\r$ represent steeper $T(P)$ profiles, which generally intersect the immiscibility gap over a more narrow range in pressure. This can be seen by comparing the two solar-age profiles in the right panel of Figure~\ref{fig.phase_diagram}, and is manifested in the relative widths of the features in $Y$ and $\grad$ in Figure~\ref{fig.he4_grad_lum_profiles}.

\begin{figure}[ht]
	\begin{center}
		\includegraphics[width=\columnwidth]{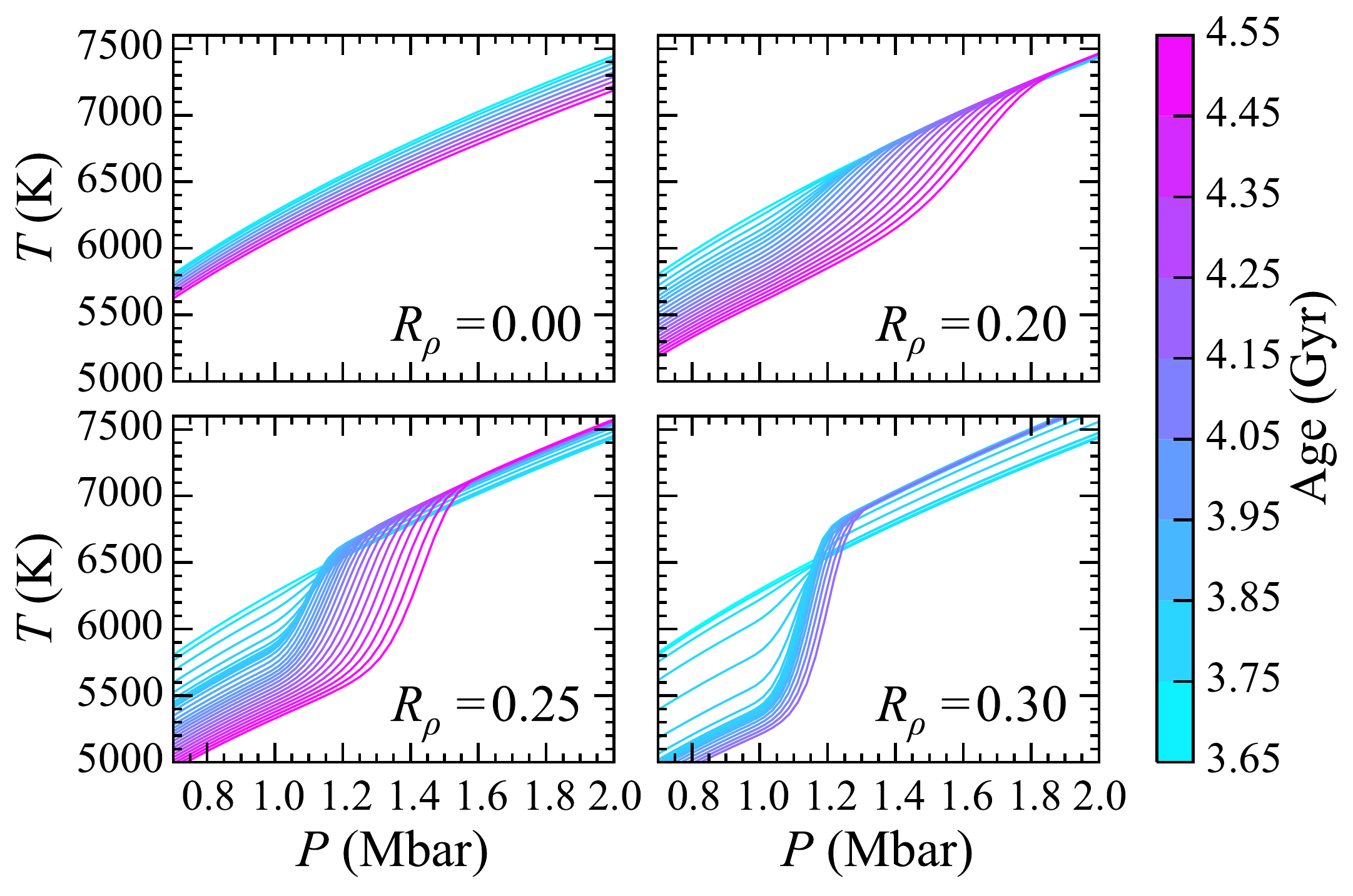}
		\caption{Evolution of the interior temperature profile for differentiating Jupiter models (the same models as in Figure~\ref{fig.he4_grad_lum_profiles}). Each panel plots a time sequence of profiles for the model indicated. Color maps to model age.}
		\label{fig.late_t_profiles}
	\end{center}
\end{figure}

\begin{figure}[ht]
	\begin{center}
		\includegraphics[width=0.9\columnwidth]{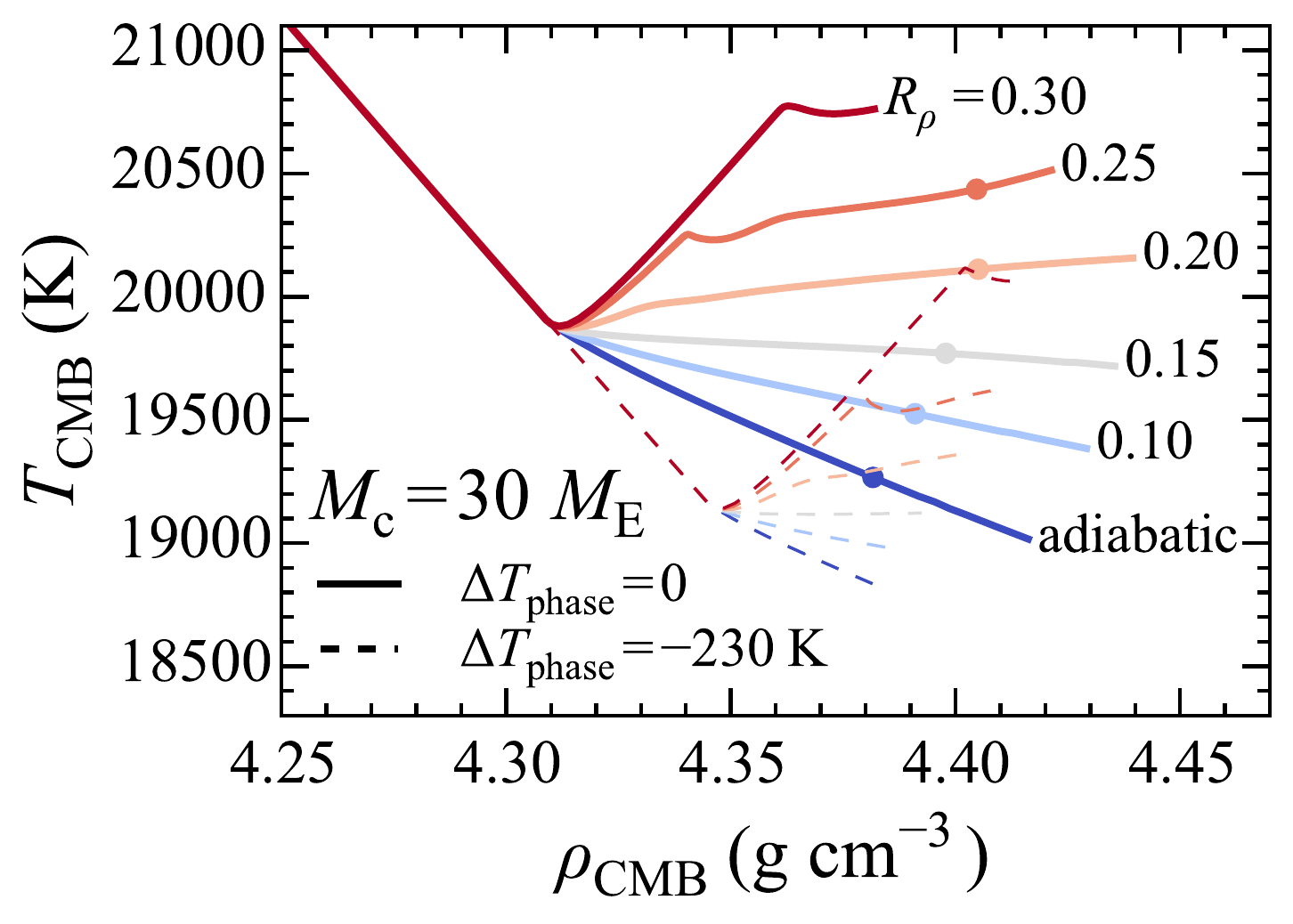}
		\caption{Evolution of the temperature and density at the core-mantle boundary (the innermost grid point in our simulations) for six different values of the superadiabaticity $\r$ of the temperature profile in the helium gradient region. The two families of curves are for two representative temperature offsets applied to the H-He phase diagram: the solid tracks assume $\dtdmx=0$; the dashed tracks assume $\dtdmx=-230$ K. For each model with $\dtdmx=0$ that reaches the solar age, that point is indicated with a filled circle.}
		\label{fig.tc_rhoc_superad}
	\end{center}
\end{figure}

The evolution is more complex in the case of nonzero superadiabaticity $\r>0$, owing to the feedback between the composition profile and the temperature profile. Since in this case the settling of helium directly modifies $T(P)$ via Equation~\ref{eq.r0}, it contributes to both the thermal (Equation~\ref{eq.energy_PT}) and composition (Equation~\ref{eq.eps_grav_comp}) components of $dL/dm$. For cases with $\r\gtrsim0.20$, double-diffusive convection poses a sufficiently strong thermal barrier that the homogeneous, adiabatic deep interior actually heats up with time, and the reduced heat flux impinging on the bottom of the molecular envelope allows the envelope to cool relatively quickly. The temperature evolution in the vicinity of the helium gradient is illustrated for these same four values of $\r$ in Figure~\ref{fig.late_t_profiles}, and Figure~\ref{fig.tc_rhoc_superad} shows the evolution of the core-mantle boundary in $T-\rho$ space. Indeed since a more pronounced temperature contrast over the helium gradient region drives a steeper composition gradient as per the phase diagram, a runaway effect ensues, with $\teff$ and $\yatm$ plummeting as stronger stratifications are realized.  In the most extreme case shown here ($\r=0.30$), the effective temperature decreases by 8 K over $10^8$ yr in this phase, to be compared to the roughly 1 K per $10^8$ yr cooling rate before the onset of helium rain. After roughly a thermal time for the homogeneous interior, the gradient region feels significant heating from below, quenching the runaway and once again assuming a state of slow evolution in which the surface cools by roughly 1 K per Gyr.

\begin{figure}[ht]
	\begin{center}
		\includegraphics[width=0.9\columnwidth]{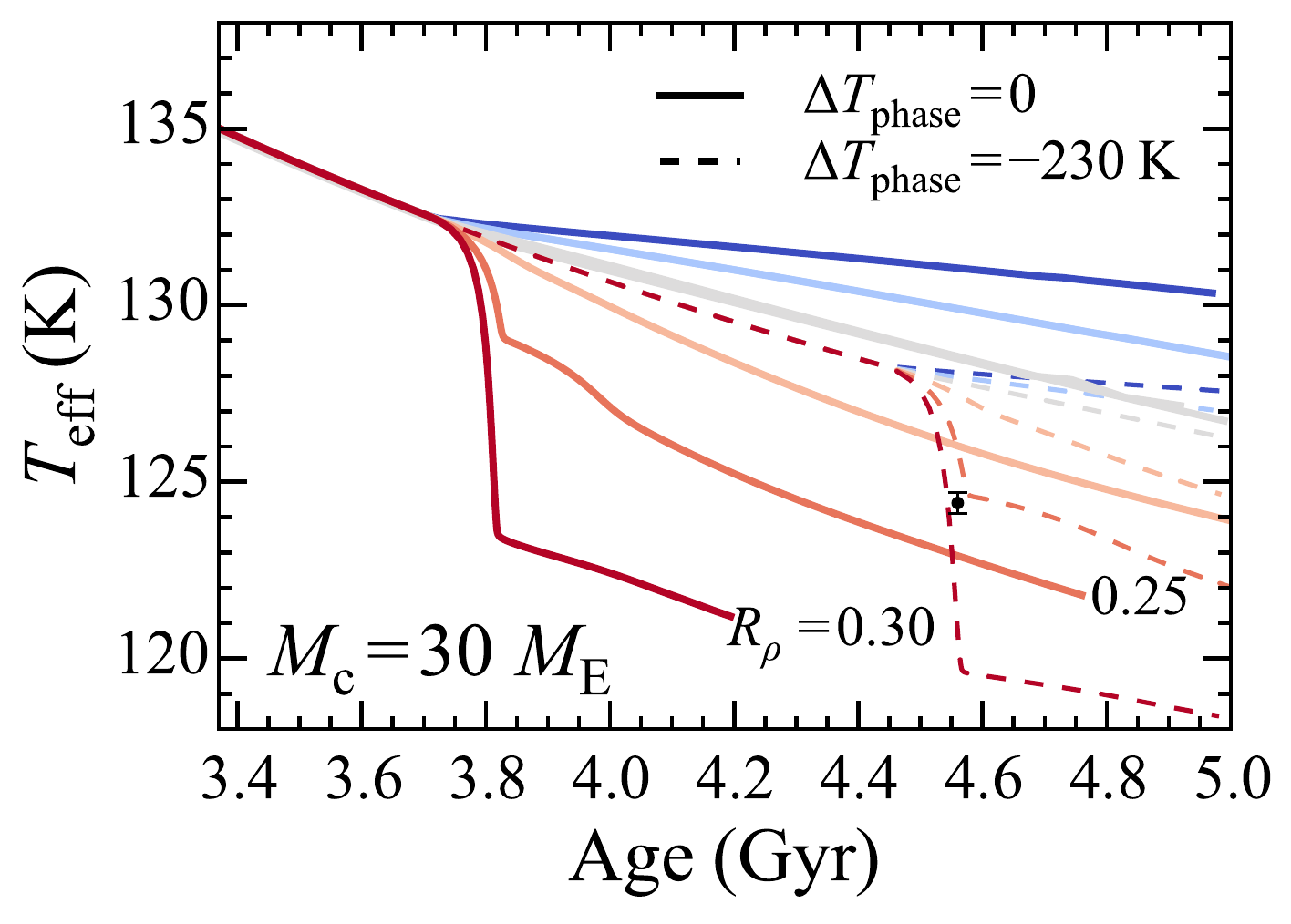}
		\caption{As in Figure~\ref{fig.tc_rhoc_superad}, but showing the effective temperatures as a function of age. The marker shows Jupiter's observed $\teff$ at the solar age.}
		\label{fig.teff_age_superad}	
	\end{center}
\end{figure}

The effect of translating the phase curve in temperature is summarized in Figures~\ref{fig.tc_rhoc_superad} and \ref{fig.teff_age_superad}, which show evolutionary tracks for two families of models, one with the phase diagram unmodified (solid curves) and one with a representative offset of $\dtdmx=-230$ K for illustration purposes (dashed curves). The crossing of the two families of curves in Figure~\ref{fig.teff_age_superad} demonstrates the fundamental anticorrelation between $\r$ and $\dtdmx$. For instance, an effective temperature of $124$ K at $4.56\ {\rm Gyr}$ can be realized either by a model with relatively modest superadiabaticity using the unmodified phase diagram, or by a model with a more extreme superadiabaticity and a delayed helium rain onset. However, the offset of $\dtdmx=-230$ K delays the onset of helium rain by nearly 800 Myr and consequently leads to a more modest depletion of helium from the molecular envelope by the time the model reaches the solar age. As demonstrated by \cite{2015MNRAS.447.3422N} and discussed in \S\ref{s.models} above, a downward offset of roughly this magnitude must be applied to the \cite{2011PhRvB..84w5109L} phase diagram to yield values of $\yatm$ at the solar age which are consistent with the Galileo entry probe measurement (Table~\ref{t.data}). Lastly, we note that translating the phase diagram to lower temperatures also leads to a more localized helium gradient region, the $\dtdmx=-230$ K case yielding a gradient roughly 1/3 the geometric thickness of the gradient established in the $\dtdmx=0$ case for each of the $\r$ values considered here. In the following section, we leave $\dtdmx$ as a free parameter alongside $\mc$ and $\r$ and systematically estimate all three simultaneously by fitting Jupiter's $\teff$ and $\yatm$ along with its volumetric radius $\rvol$.

\subsection{Bayesian Parameter Estimation}\label{s.mcmc}
We estimate model parameters and their statistical uncertainties using Markov chain Monte Carlo (MCMC).  In particular, given our nonlinear three-parameter model
\begin{equation}
	\theta\equiv\{\mc,\ \r,\ \dtdmx\}
\end{equation} 
and fundamental Jupiter data (see Table~\ref{t.data})
\begin{equation}
	 D\equiv\{\teff,\ \yatm,\ \rvol\},
\end{equation}
we calculate the posterior probability distribution from Bayes' theorem
\begin{equation}
	P(\theta| D)\propto P( D|\theta)P(\theta).\label{eq.bayes}
\end{equation}
In the likelihood $P(D|\theta)$ we assume Gaussian errors for the data $D$:
\begin{align}\label{eq.likelihood}
\ln P( D|\theta)=-\frac12\Big[
& \ln(2\pi\sigma_{\teff}^2) + \frac{(\teff-\teff^{\rm m})^2}{\sigma_{\teff}^2} \nonumber\\
&+\ln(2\pi\sigma_{\yatm}^2) + \frac{(\yatm-\yatm^{\rm m})^2}{\sigma_{\yatm}^2} \nonumber\\
&+\ln(2\pi\sigma_{\rvol}^2) + \frac{(\rvol-R^{\rm m})^2}{\sigma_{\rvol}^2}\Big],
\end{align}
where a superscript m denotes the model outcome at the solar age; each sample from the posterior distribution thus entails a full evolutionary calculation.  Samples are drawn from the posterior distribution Equation~\ref{eq.bayes} using the Python MCMC implementation \emcee \citep{2013PSP..125..306F}. Our prior probability distribution $P(\theta)$ is chosen to be uniform over $\mc\geq0$, all $\dtdmx$, and $0.1\leq \r\leq1$ following the criteria for linear instability (see \S\ref{s.semiconvection_model}, in particular Equation~\ref{eq.rcrit}); elsewhere it is zero. 

\begin{deluxetable}{ccccc} 
\tabletypesize{\footnotesize} 
\tablecolumns{5} 
\tablewidth{0pt} 
\tablecaption{Global Jupiter Data \label{t.data}} 
\tablehead{ 
\colhead{Age (yr)} &
\colhead{$\teq$ (K)} &
\colhead{$\teff$ (K)} & 
\colhead{$\yatm$} & 
\colhead{$\rvol$ (km)} 
} 
\startdata 
$4.56\times10^9$ &
$109.0$\tablenotemark{a} &
$124.4\pm0.3$\tablenotemark{b} & 
$0.234\pm0.005$\tablenotemark{c} & 
$69911\pm6$\tablenotemark{d}
\enddata 
\tablenotetext{a}{\S\ref{s.atmospheres}}
\tablenotetext{b}{\cite{1981Sci...212..192H}}
\tablenotetext{c}{\cite{1998JGR...10322815V}}
\tablenotetext{c}{\cite{2007CeMDA..98..155S}}
\end{deluxetable}

\begin{figure}[ht]
	\begin{center}
		\includegraphics[width=\columnwidth]{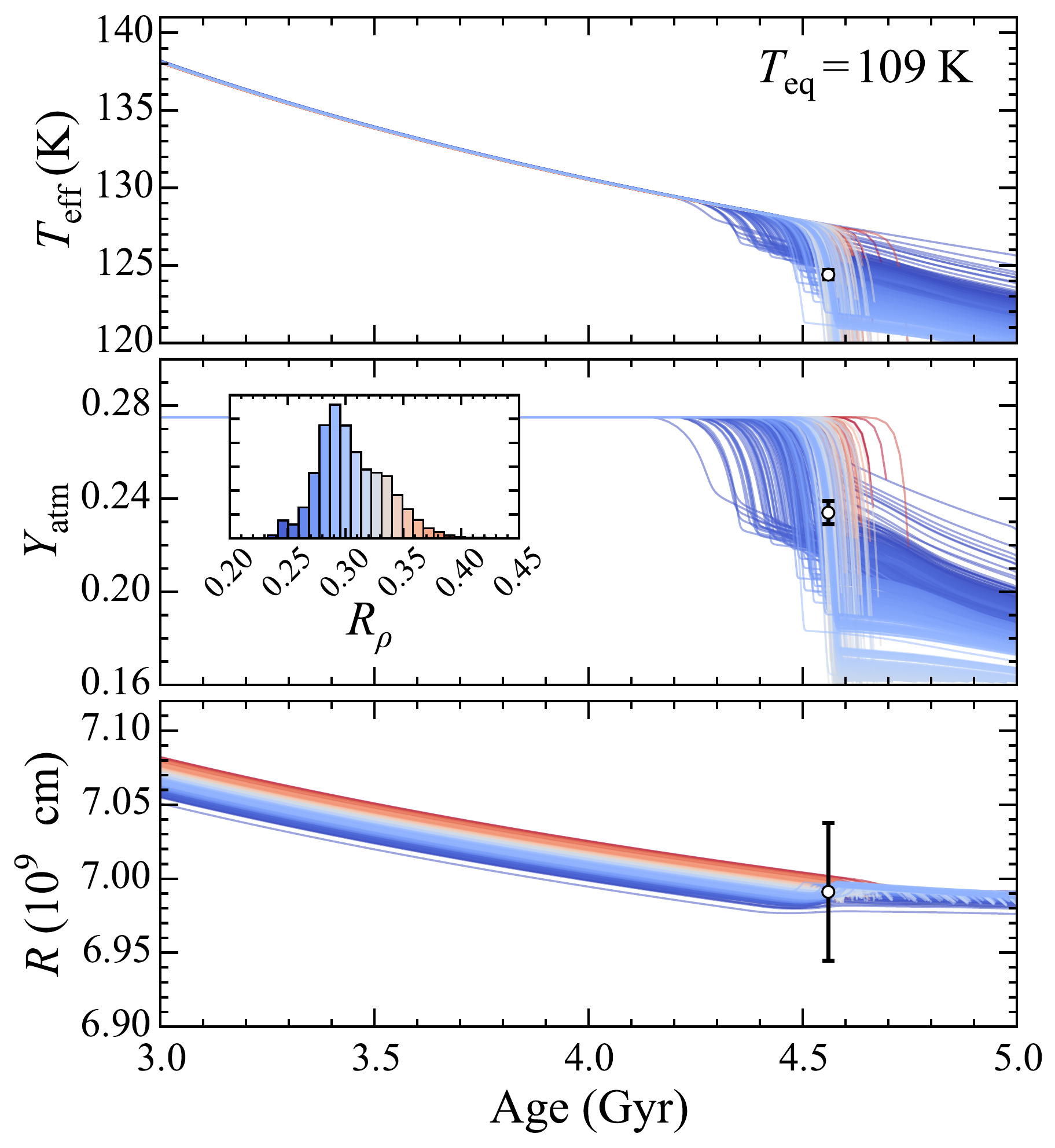}
		\caption{Late evolution of Jupiter models undergoing helium rain and overstable double-diffusive convection, sampled using MCMC. Open circles with error bars designate observations (Table~\ref{t.data}); the error on $\teff$ is smaller than the marker. The color of an evolutionary a track encodes its $\r$ value as specified by the colors of the bars in the histogram (inset, center panel), which is a coarsely binned version of the marginalized posterior probability distribution in Figure~\ref{fig.ndim3_posteriors}. Only a random subset of tracks are shown, and more likely tracks are plotted on top of less likely ones.}
		\label{fig.ndim3_tracks}
	\end{center}
\end{figure}

The late evolution of $\teff$, $\yatm$ and $R$ for the a subset of the evolutionary sequences in the resulting chain are shown in Figure~\ref{fig.ndim3_tracks}, which color codes tracks by their value of $\r$. For the duration of the initial homogeneous phase, the differences in $\teff$ and $R$ between tracks stem solely from the heavy element core mass $\mc$, since that parameter adjusts the mean density and hence the total radius of the planet. As described in \S\ref{s.inhomog}, the time at which phase separation sets in (and $\yatm$ first diverges from the protosolar value) is set by $\dtdmx$, and the trend toward later phase separation onset with increasing $\r$ is the result of the two parameters' substantial covariance. Timesteps spanning the solar age are typically $10^7$ yr, and the values of $\teff$, $\yatm$ and $R$ were interpolated linearly within those timesteps to obtain $\teff^{\rm m}$, $\yatm^{\rm m}$, and $R^{\rm m}$. Proposed steps in which the model terminated before the solar age (as would be expected for a model with simultaneously large $\r$ and $\dtdmx$, for example) were rejected, since a calculation of the likelihood (Equation~\ref{eq.likelihood}) would not be possible.

The models that terminate before 5 Gyr did so either because (i) the model cooled to $\teff=120$ K, at which point we stop to avoid unnecessary computation time, or (ii) upon the onset of phase separation, the luminosity inversion described in the previous subsection (and evident in the $\r=0.30$ case in Figure~\ref{fig.he4_grad_lum_profiles}) grew to the degree that a negative luminosity was realized in the interior, typically just outside the helium gradient region. This behavior follows from attempting to enforce large values of $\r$ such that the the large luminosity generated by deposition of helium into the metallic interior cannot be communicated upward through the weakly turbulent double-diffusive layer. \cite{2015MNRAS.447.3422N} identified the same effect in their models with LDDC, noting that there exists a minimum layer height such that the luminosity is still positive throughout the interior. In our models this translates to an effective upper limit on the values of $\r$ attainable by models with strictly positive luminosity profiles. 

\begin{figure}[ht]
	\begin{center}
		\includegraphics[width=\columnwidth]{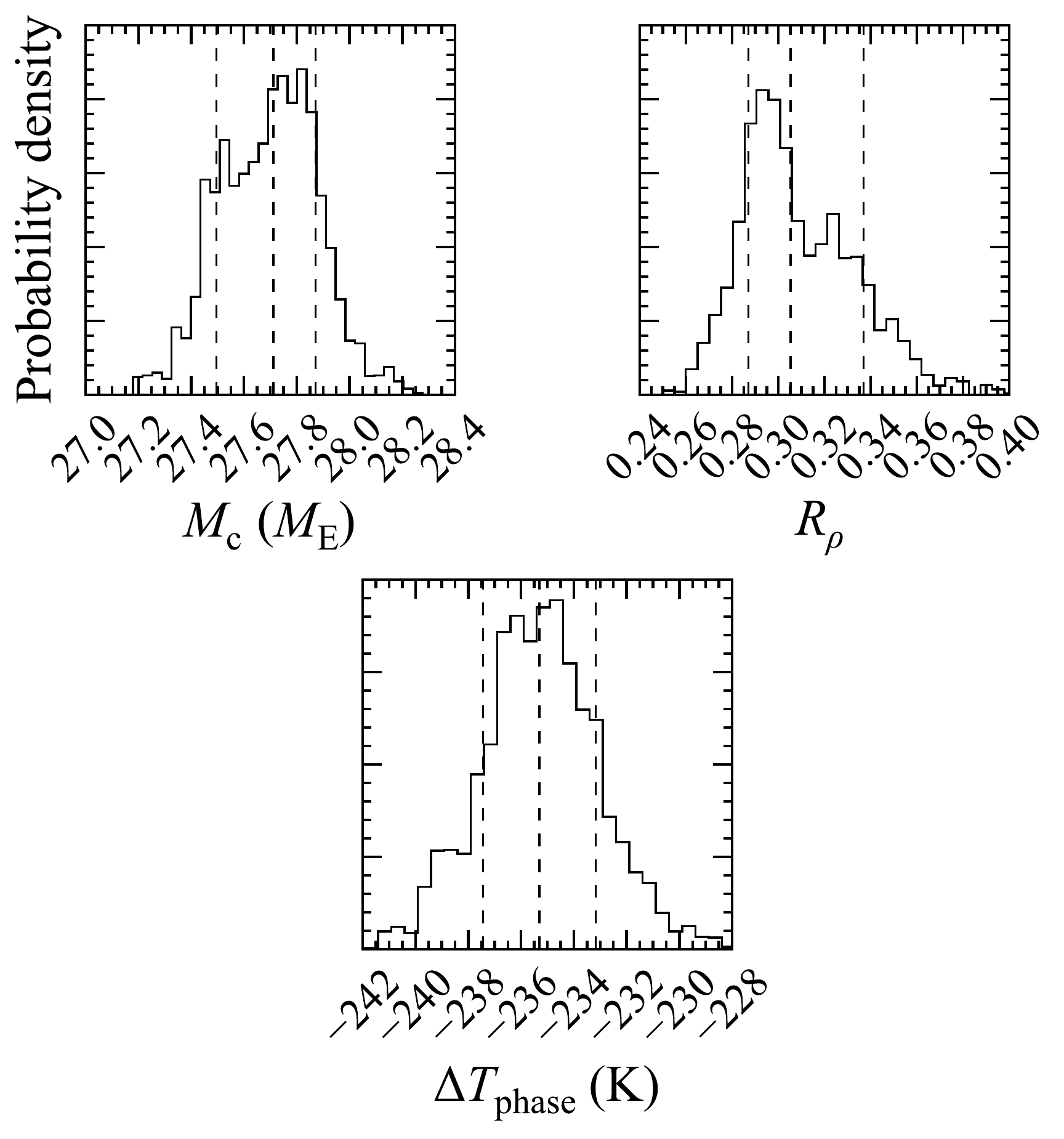}
		\caption{Posterior probability distributions for the heavy element mass $\mc$, density ratio $\r$ in the helium gradient region, and phase diagram offset $\dtdmx$ based on the evolutionary sequences shown in Figure~\ref{fig.ndim3_tracks}. Each distribution is the full three-dimensional joint distribution marginalized over the other two parameters. The vertical dashed line near each peak designates the median, with the flanking vertical dashed lines enclosing the central 68\% of cumulative probability.}
		\label{fig.ndim3_posteriors}
	\end{center}
\end{figure}

The outcome of our Bayesian parameter estimation is the posterior probability distribution shown in Figure~\ref{fig.ndim3_posteriors}, wherein each panel plots the full joint distribution marginalized over the other two parameters. The medians of each distribution are indicated, as are the central 68\% confidence regions.  Typical models (as characterized by the medians) have massive cores ($\mc=27.7\ \me$), are strongly superadiabatic in the compositionally stratified region ($\r=0.31$), and have substantial downward offsets to the phase diagram ($\dtdmx=-235$ K.) The posterior distribution of $\dtdmx$ is narrowly peaked at these large negative offsets, and quite robustly rules out the unmodified phase diagram.  This result is driven by the requirement that the model's homogeneous molecular envelope retains enough helium to match the modest depletion measured by the Galileo entry probe (Table~\ref{t.data}).

The single largest modeling uncertainty for the cooling of the giant planets is associated with the equation of state. Depending on the assumed EOS, \cite{2004ApJ...609.1170S} found that cooling times for homogeneous Jupiter models spanned 3.1 to 5.4 Gyr. (For comparison, our homogeneous models with \scvhi cool in 5 to 5.5 Gyr for realistic heavy element masses and equilibrium temperatures; see Figures~\ref{fig.homog_tracks} and \ref{fig.homog_cooling_times}.) The range in cooling times obtained from applying different equations of state owes mostly to the fact that for different $P(T)$ relations, models with a given entropy possess different total thermal energy content and hence take more or less time to cool to Jupiter's present-day luminosity.

As a means of exploring how our results would be affected by the application of a different equation of state, we repeat our full calculations with a modified atmospheric boundary condition. Although the model atmosphere and equation of state are not directly related, they are degenerate in that they both dictate the overall timescale for the thermal evolution. This can be made explicit by first integrating the energy equation (equation \ref{eq.tds}) over the mass of the planet to yield
\begin{equation}
	L_{\rm int}=-\int_0^M T\frac{ds}{dt}\,dm=4\pi R^2\sigma_{\rm SB}\!\left(T_{\rm eff}^4-T_{\rm eq}^4\right).\label{eq.int_tdsdt_dm}
\end{equation}
For the simplified example of a planet cooling through a sequence of isentropes, $ds/dt$ is independent of $m$ and the second equality can be integrated to yield the total cooling time
\begin{equation}
	\tau_{\rm cool}=\int_0^{\tau_{\rm cool}} dt=\int_{s_{\rm cool}}^{s_0}\left(\frac{\int_0^MT(m, s)\,dm}{4\pi R^2(s)\sigma_{\rm SB}\!\left(\teff^4(s)-\teq^4\right)}\right)\,ds,\label{eq.tau_cool}
\end{equation}
where $s_0$ designates an arbitrary large starting entropy, $s_{\rm cool}$ designates the planet's current entropy, and other symbols have their usual meanings. Equation~\ref{eq.tau_cool}, while emphatically not how our evolutionary sequences are calculated, serves as a heuristic tool to demonstrate that choosing a colder EOS (such that the mean temperature along a given adiabat is lower) and reducing the solar input $\teq$ have a similar effect.

Our modification of the boundary condition as a proxy for a different H-He EOS is motivated by the lack of other realistic EOS options presently available in \mesa at the relevant densities and temperatures. Varying the parameter $\teq$ offers a simple means of producing a different total cooling time, the relation between the two being illustrated in Figure~\ref{fig.homog_cooling_times}. As an example, we find that a homogeneous, adiabatic model with $\mc=30\ \me$ and $\teq$ reduced to 100 K cools to Jupiter's $\teff$ in just 4.2 Gyr. Since very large superadiabaticities tend to reduce the cooling (see Figure~\ref{fig.teff_age_superad}), differentiating models that satisfy the basic constraints of Table~\ref{t.data} in spite of a cold boundary condition must have small values for $\r$ such that the cooling is extended to the age of the solar system. 

\begin{figure}[ht]
	\begin{center}
		\includegraphics[width=\columnwidth]{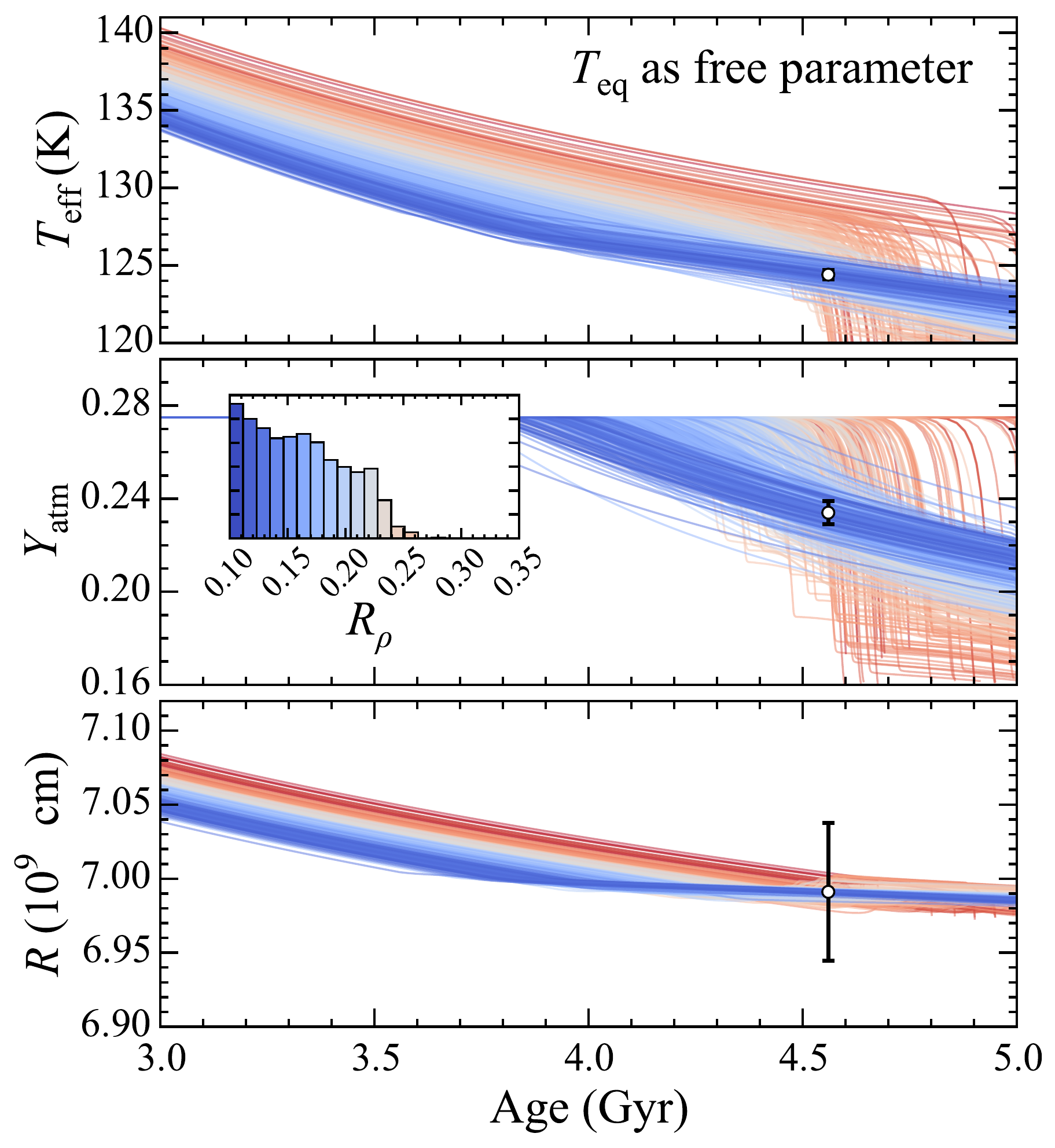}
		\caption{As in Figure~\ref{fig.ndim3_tracks}, but including $\teq$ as a additional free parameter to mimic the effect of an EOS predicting a warmer or colder interior.}
		\label{fig.ndim4_tracks}
	\end{center}
\end{figure}

\begin{figure}[ht]
	\begin{center}
		\includegraphics[width=\columnwidth]{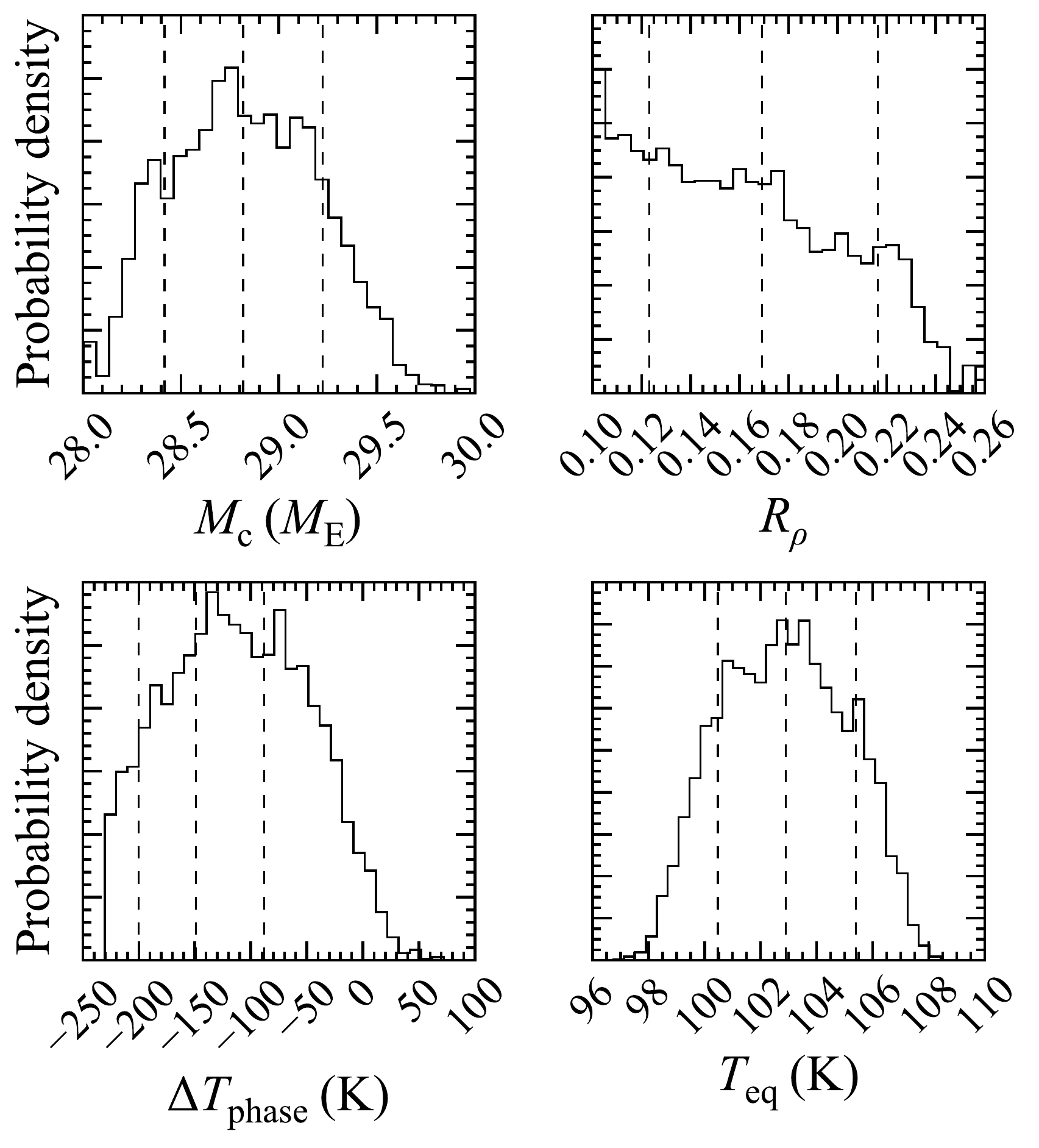}
		\caption{As in Figure~\ref{fig.ndim3_posteriors}, but including $\teq$ as an additional free parameter to mimic the effect of an EOS predicting a warmer or colder interior.}
		\label{fig.ndim4_posteriors}
	\end{center}
\end{figure}

Figure~\ref{fig.ndim4_tracks} shows the late evolution of models computed with this artificially free boundary condition, with a uniform prior chosen for $\teq$; the posterior distributions of $\mc$, $\r$, $\dtdmx$ and $\teq$ are shown in Figure~\ref{fig.ndim4_posteriors}. In this case the extra freedom in the boundary condition leads to a wide variety of total times spent in the homogeneous phase of evolution, and consequently the other parameters $\mc$, $\r$ and $\dtdmx$ have markedly wider posterior probability distributions. Most likely models have equilibrium temperatures several K cooler than the measured value $\teq=109\ {\rm K}$ such that the homogeneous cooling is more short-lived, and in the inhomogeneous evolution that follows, double-diffusive convection can proceed with temperature gradients closer to the adiabat and still satisfy the $\teff$ constraint at the solar age. Indeed, the models display an overall preference for the lowest possible density ratios; the marginalized posterior probability density increases uniformly toward lower values of $\r$ and peaks at the lower boundary imposed by the step-function prior $\r>\rcrit=0.1$, which was imposed in light of the linear criterion for the double-diffusive instability (see \S\ref{s.ddc_param}). As a result, the most likely evolution during the inhomogeneous phase is secular cooling of the envelope, with no dropoff of the surface temperature or helium abundance over short timescales. These models might be considered preferable to the most likely models obtained in the three-parameter case of Figures~\ref{fig.ndim3_tracks} and \ref{fig.ndim3_posteriors} because if $\teff$, $\yatm$, and $\rvol$ were undergoing drastic changes on a $10^8$ year timescale, then observing Jupiter in its present state would be somewhat serendipitous.

Freeing $\teq$ also allows much more modest corrections to the phase diagram, the 68\% credible interval now spanning $-200$ to $-100$ K. Importantly, despite the freedom in the overall cooling time, the unmodified phase diagram is still found to be extremely unlikely with 95\% of probability lying at offsets $\dtdmx<-50\ {\rm K}$. 

The posterior distribution of $\teq$ values in Figure~\ref{fig.ndim4_posteriors} should not be interpreted as a new determination of Jupiter's $\teq$, as that is a measured quantity. Rather, it contains information about a most likely equation of state: it is significant that the distribution excludes the measured value almost completely, with only 0.4\% of cumulative probability within the error of the measurement (Table~\ref{t.data}). If we suppose that our model contains the essential physics, this can be taken as evidence that the real EOS for hydrogen and helium predicts substantially colder interiors than does \scvhi. For reference, a homogeneous evolutionary sequence computed with the median values $\mc=28.8\ \me$ and $\teq=102.9\ {\rm K}$ from the distributions in Figure~\ref{fig.ndim4_posteriors} cools to Jupiter's $\teff$ in 4.46 Gyr (see Figure~\ref{fig.homog_cooling_times}).

\section{Discussion}
The framework developed here consists of a Python class for creating instances of \mesa work directories, modifying \mesa inlists, executing the evolution program, and processing its output, all as part of a single likelihood calculation called by an \emcee sampler. The method renders it straightforward to add additional parameters to the model or incorporate new or updated constraints in any quantity output by \mesa directly. It is readily adaptable to a host of different problems, most obviously non-adiabatic thermal evolution models for Saturn, where the same fundamental physics operate. Beyond just the Jovian planets and H-He immiscibility, our technique has broad applicability for deriving properties of objects from giant planets to brown dwarfs and stars, e.g., retrieving the age and composition for an object with measured mass and radius. Performing these retrievals with a code as mature as \mesa means that our knowledge of stellar/planetary evolution is built in, including complexities such as self-consistently determining mixing boundaries, modeling double-diffusive transport processes, or calculating nuclear energy generation rates with state of the art nuclear networks. Thus in the example of retrieving an object's composition and age from its measured mass and radius, meaningful inferences can be made about not just the bulk composition, but the composition profile, and indeed the composition profile's possible origins and evolution. The Bayesian approach automatically provides meaningful error bars for model parameters, and combining it with the open source \mesa package offers more flexibility than traditional grid-based isochrone fitting because new parameters---and indeed new physics---can be added at will. 

This work builds on that of \cite{2015MNRAS.447.3422N} principally in two ways: first, it makes the weakest possible assumption about the temperature gradient resulting from double-diffusive convection in the deep interior, abandoning the assumption of layered convection following the flux laws derived by \cite{2013ApJ...768..157W} in favor of a generic model wherein any temperature gradient can be attained as long as it is consistent with the criterion for linearly overstable gravity waves. Second, performing the calculations in an MCMC framework allows a probabilistic determination of all model parameters simultaneously, and we find a multitude of models that satisfy the imposed constraints (Table~\ref{t.data}).  We demonstrated that \scvhi predicts strongly superadiabatic temperature profiles in Jupiter's helium gradient region, such that the planet's surface cools rapidly as most of the metallic hydrogen interior heats up over time. Repeating the calculations with a variable boundary condition to probe the effects of using a different EOS, we found that more modest superadiabaticities are preferred, although the distribution of allowed values is still broad. We found in all cases that the unperturbed phase diagram of \cite{2011PhRvB..84w5109L} is highly unlikely.

That such a diversity of models meeting the imposed constraints were obtained in Figures~\ref{fig.ndim4_tracks} and \ref{fig.ndim4_posteriors} underscores the severe uncertainties that persist in modelling the evolution of giant planets. Admittedly, the present work does not exploit all the available data. Most importantly, our models make no use of of Jupiter's gravitational harmonics or its axial moment of inertia, both of which constrain the interior density profile. As discussed in \S\ref{s.models}, a calculation of oblateness and the associated non-spherical components of the gravity field is beyond our scope because the only EOS currently available for modeling giant planets in \mesa, \scvhi, is limited to hydrogen and helium. All heavy elements are in an inert core rather than partially distributed through the envelope, and as such the density profiles in our models are somewhat unrealistic and are not suited for fitting to $J_2$ or any higher-order moments. Nonetheless, we view these models as complimentary to the detailed static models computing using more realistic equations of state (e.g., \citealt{hubbard2016}) in that we use a forward thermal evolution model to derive estimates for Jupiter's deep superadiabatic temperature stratification and corrections to the H-He phase diagram, both of which should be taken into consideration for improving static models of the Jovian planets. Our findings support the existing body of evidence indicating that a realistic H-He equation of state departs significantly from \scvhi.

\acknowledgements
We thank Nadine Nettelmann for thoughtful comments that helped to significantly improve the manuscript. C.M. acknowledges support from NASA Headquarters under the NASA Earth and Space Science Fellowship Program (grant NNX15AQ62H). Simulations for this research were carried out on the UCSC supercomputer Hyades, which is supported by National Science Foundation (grant AST-1229745) and the University of California, Santa Cruz.



\end{document}